\documentclass[12pt, draftclsnofoot, onecolumn]{IEEEtran}
\usepackage{amssymb}
\usepackage{amsmath}
\usepackage{cite}
\usepackage{url}
\usepackage{xcolor}
\usepackage{cite,graphicx,amsmath,amssymb}
\usepackage{subfigure}
\usepackage{citesort}
\usepackage{fancyhdr}
\usepackage{mdwmath}
\usepackage{mdwtab}
\usepackage{caption}
\usepackage{amsthm}
\usepackage{setspace}
\usepackage{algorithm}
\usepackage{algorithmic}
\usepackage{makecell}
\usepackage{diagbox}
\newcommand{\bm}[1]{\mbox{\boldmath{$#1$}}}
\newcommand{\bmm}[1]{\mbox{\footnotesize \boldmath{$#1$}}}

\newtheorem{remark}{Remark}
\newtheorem{theorem}{Theorem}

\newtheorem{lemma}{Lemma}

\newtheorem{corollary}{Corollary}

\allowdisplaybreaks
\makeatletter
\def\ScaleIfNeeded{%
\ifdim\Gin@nat@width>\linewidth \linewidth \else \Gin@nat@width
\fi } \makeatother
\begin{document}
\title{Simultaneously Transmitting And Reflecting (STAR) RIS Aided Wireless Communications}
% \markboth{\textit{A Manuscript Submitted to The IEEE   Communications Magazine} }: Operating Protocols and Joint Beamforming Design

\author{

Xidong~Mu,~\IEEEmembership{Graduate Student Member,~IEEE,}
        Yuanwei~Liu,~\IEEEmembership{Senior Member,~IEEE,}
       Li~Guo,~\IEEEmembership{Member,~IEEE,}
       Jiaru~Lin,~\IEEEmembership{Member,~IEEE,}
       and Robert~Schober,~\IEEEmembership{Fellow,~IEEE}

\thanks{Part of this work has been submitted to the Asilomar Conference on Signals, Systems, and Computers, Oct. 31-Nov. 3, 2021~\cite{Asilomar_xidong}.}
\thanks{X. Mu, L. Guo, and J. Lin are with the Key Laboratory of Universal Wireless Communications, Ministry of Education, Beijing University of Posts and Telecommunications, Beijing, China, and are also with the School of Artificial Intelligence, Beijing University of Posts and Telecommunications, Beijing, China. (email:\{muxidong, guoli, jrlin\}@bupt.edu.cn).}
\thanks{Y. Liu is with the School of Electronic Engineering and Computer Science, Queen Mary University of London, London, UK. (email:yuanwei.liu@qmul.ac.uk).}
\thanks{R. Schober is with the Institute for Digital Communications, Friedrich-Alexander-University Erlangen-N{\"u}rnberg (FAU), Germany (e-mail: robert.schober@fau.de).}
}

%, namely transmission/reflection beamforming design and research allocation, Compared with conventional reflecting-only RISs,
%, where a base station (BS) sends information via the transmission and reflection links to users surrounding a STAR-RIS.
%For multicast communication, the BS sends common messages to the two users. In this case, the power consumption minimization problem for joint beamfroming design is formulated to satisfy the system multicasting rate requirement. The proposed algorithms for each protocol in unicast are shown to be applicable to the problems in multicast with some modifications.
\maketitle
\vspace{-1.5cm}
\begin{abstract}
The novel concept of simultaneously transmitting and reflecting (STAR) reconfigurable intelligent surfaces (RISs) is investigated, where the incident wireless signal is divided into transmitted and reflected signals passing into both sides of the space surrounding the surface, thus facilitating a \emph{full-space} manipulation of signal propagation. Based on the introduced basic signal model of `STAR', three practical operating protocols for STAR-RISs are proposed, namely energy splitting (ES), mode switching (MS), and time switching (TS). Moreover, a STAR-RIS aided downlink communication system is considered for both unicast and multicast transmission, where a multi-antenna base station (BS) sends information to two users, i.e., one on each side of the STAR-RIS. A power consumption minimization problem for the joint optimization of the active beamforming at the BS and the passive transmission and reflection beamforming at the STAR-RIS is formulated for each of the proposed operating protocols, subject to communication rate constraints of the users. For ES, the resulting highly-coupled non-convex optimization problem is solved by an iterative algorithm, which exploits the penalty method and successive convex approximation. Then, the proposed penalty-based iterative algorithm is extended to solve the mixed-integer non-convex optimization problem for MS. For TS, the optimization problem is decomposed into two subproblems, which can be consecutively solved using state-of-the-art algorithms and convex optimization techniques. Finally, our numerical results reveal that: 1) the TS and ES operating protocols are generally preferable for unicast and multicast transmission, respectively; and 2) the required power consumption for both scenarios is significantly reduced by employing the proposed STAR-RIS instead of conventional reflecting/transmiting-only RISs.
\end{abstract}
%\begin{keywords}
%Operating protocols, reconfigurable intelligent surfaces, simultaneous transmission and reflection, unicast and multicast communication.
%\end{keywords}
\section{Introduction}%increase the numbers of antennas and base stations (BSs), i.e.,  and explore higher radio frequencies, i.e.,
With the worldwide commercialization of fifth-generation (5G) wireless communication networks, growing research efforts are being devoted to the upcoming beyond 5G (B5G) and sixth-generation (6G) wireless communication networks, which are expected to impose more stringent requirements such as extremely high spectrum- and energy-efficiency, microsecond latency, and full-dimensional network coverage~\cite{Saad_6G,Zhang_6G}. To achieve these goals, extensions to the existing communication technologies~\cite{5G}, which have already been proven effective, have been proposed, including but not limited to ultra-massive multiple-input multiple-output (UM-MIMO), ultra-dense networks (UDNs), and terahertz (THz) communication. However, on the one hand, the relentless increase of the number of antennas/base stations (BSs) and the use of very high carrier frequencies will potentially cause high energy consumption and hardware cost, since more power-hungry and costly radio frequency (RF) chains have to be installed for signal conversion. On the other hand, deploying such a large number of active components operating at very high frequencies in wireless networks may not always be beneficial since they also introduce new challenges, such as complicated inter-user/cell interference scenarios, pilot contamination, and severe hardware impairments.\\
\indent To overcome the above limitations, new cost-effective techniques have to be developed for wireless communication systems. Specifically, motivated by the rapid development of metasurfaces and corresponding advanced fabrication technologies, reconfigurable intelligent surfaces (RISs) have emerged as promising solutions~\cite{Basar,Renzo_IRS,RIS_survey,WuTowards,Huang_Mag}. A RIS is a man-made two-dimensional (2D) surface, which is equipped with a large number of low-cost passive elements. With the aid of a smart controller attached to the RIS, the propagation of the wireless signals incident on the RIS can be adjusted through the customized phase response of each element. Therefore, RIS can facilitate ``Smart Radio Environments (SREs)''~\cite{Renzo_IRS}. Compared to the conventional multi-antenna and relaying concepts, where wireless signals are actively \emph{produced} using costly RF chains, RISs only passively \emph{recycle} signals that are already available in the network and do not require RF chains. Thus, RISs are more economical and environmentally friendly compared to conventional active antenna systems. Furthermore, RISs can be seamlessly integrated into existing wireless networks by deploying them on diverse structures, such as roadside billboards, building facades, windows, and even human clothes~\cite{RIS_survey}.
\vspace{-0.4cm}
\subsection{Prior Works}%from both industry and academy
\vspace{-0.2cm}
Motivated by the aforementioned favorable characteristics of RISs, extensive research efforts have been devoted to exploiting the new degrees-of-freedom (DoFs) introduced by RISs with the prospect of mitigating a wide range of challenges encountered in wireless networks, such as the reduction of transmit power consumption~\cite{Wu2019IRS,Yu_optimal,Zheng_NOMA}, the improvement of spectrum- and energy-efficiency~\cite{Huang_EE,Mu_IRS,Zhang,Mu_capacity,Xu_Cognitive,Huang_IRS_DL}, and the establishment of unobstructed communication links~\cite{Zheng_double_IRS,Mei}. More specifically, the authors of \cite{Wu2019IRS} proposed a suboptimal alternating optimization based algorithm to minimize the transmit power of the access point (AP) by jointly optimizing the active beamforming at the AP and the passive beamforming at the RIS. The transmit power minimization problem was further studied by the authors of \cite{Yu_optimal} for single-user systems, where a globally optimal solution was obtained with a branch-and-bound method. Furthermore, the authors of \cite{Zheng_NOMA} minimized the transmit power of RIS-aided multi-user systems employing orthogonal multiple access (OMA) and non-orthogonal multiple access (NOMA), respectively. Considering a practical power consumption model for the RIS elements, the authors of \cite{Huang_EE} formulated a joint active and passive beamforming optimization problem for the maximization of the network energy-efficiency. The joint beamforming optimization problem was further studied by the authors of \cite{Mu_IRS} for a RIS-assisted downlink multiple-input single-output (MISO) NOMA system, where the system sum rate was maximized for continuous and discrete RIS phase shifters, respectively. Furthermore, the communication capacity of a RIS-aided MIMO system and a RIS-aided multi-user system was characterized in~\cite{Zhang} and~\cite{Mu_capacity}, respectively. Due to the nearly passive nature of RISs, the accurate acquisition of channel state information (CSI) is a challenging task. Hence, the authors of \cite{Xu_Cognitive} investigated robust resource allocation for a RIS-assisted full-duplex cognitive radio system with imperfect CSI. The authors of \cite{Huang_IRS_DL} invoked deep reinforcement learning to optimize the active and passive beamforming for maximization of the system sum rate based on partial CSI. To overcome signal blockages, the authors of \cite{Zheng_double_IRS} proposed to use two RISs to connect a BS to users in a signal dead zone, where cooperative passive beamforming among RISs was proposed. A similar problem was further studied for multiple RISs in \cite{Mei}. The significant benefits of deploying RISs in wireless networks have also been verified in the context of physical layer security~\cite{Yu_secure}, unmanned aerial vehicle systems~\cite{IRS_UAV}, and robotic communications systems \cite{Mu_robotic}.
\begin{figure}[b!]
  \centering
  \includegraphics[width=2.3in]{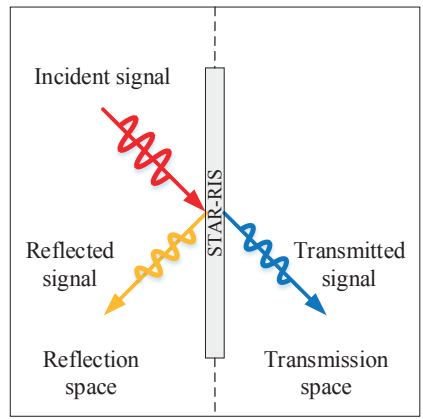}\\
  \caption{The concept of STAR-RISs }\label{STR}
\end{figure}
\vspace{-0.4cm}
\subsection{Motivations and Contributions}
\vspace{-0.2cm}
Most of the existing research contributions consider the case where the RISs are only able to reflect the incident wireless signal (referred to as conventional reflecting-only RISs). In this case, transmitter and receiver have to be located on the same side of the RIS~\cite{Wu2019IRS,Yu_optimal,Zheng_NOMA,Huang_EE,Mu_IRS,Zhang,Mu_capacity,Xu_Cognitive,Huang_IRS_DL,Zheng_double_IRS,Mei,Yu_secure,IRS_UAV,Mu_robotic}, thus leading to a \emph{half-space} SRE. However, this geographical restriction may not always be met in practice, and gravely restricts the flexibility and effectiveness of RISs, as generally users may be located on both sided of a RIS. To overcome this limitation, the novel concept of \emph{simultaneously transmitting and reflecting} RISs (STAR-RISs) was proposed in \cite{STAR,STAR_mag}. In particular, as shown in Fig. \ref{STR}, the wireless signal incident on an element of a STAR-RIS from either side of the surface is divided into two parts~\cite{tunable}. One part (reflected signal) is reflected to the same space as the incident signal, i.e., the reflection space, and the other part (transmitted signal) is transmitted to the opposite space as the incident signal, i.e., the transmission space. As shown in \cite{STAR}, by manipulating both the electric and magnetic currents of a STAR-RIS element, the transmitted and reflected signals can be reconfigured via two generally independent coefficients, namely the transmission and the reflection coefficients. Therefore, a highly flexible \emph{full-space} SRE can be realized. Two prototypes, which resemble STAR-RISs, have been developed using metasurfaces~\cite{DOCOMO,tunable}. For the prototype proposed in~\cite{tunable}, each element was composed of a parallel resonant LC tank and small metallic loops to provide the required electric and magnetic surface reactance. Despite the above advantages, the investigation of how STAR-RISs can be integrated into wireless communication systems is still in its infancy. Based on a similar idea as STAR-RISs, an intelligent omni-surface (IOS) was proposed by the authors of~\cite{IOS} to achieve full-space coverage. However, unlike for STAR-RISs, for IOSs, the phase shifts for transmission and reflection are identical, as they are determined by the state of positive intrinsic negative (PIN) diodes installed in the IOS elements. In another early work~\cite{STAR}, the authors focused on the investigation of hardware models and channel models for STAR-RISs. However, to the best of the authors' knowledge, efficient operating protocols for STAR-RISs and corresponding joint transmission and reflection beamforming optimization techniques for STAR-RIS aided wireless networks have not been studied, yet, which provides the main motivation for this work.\\
\indent To exploit the full potential of STAR-RISs, in this paper, we propose practical protocols for the operation of STAR-RISs and investigate the joint transmission and reflection beamforming design for both unicast and multicast communication. The main contributions of this paper can be summarized as follows:
\begin{itemize}
  \item We propose three practical operating protocols for STAR-RISs, namely energy splitting (ES), mode switching (MS), and time switching (TS), along with their respective benefits and drawbacks.
  \item We consider a STAR-RIS aided downlink communication system, where a BS sends information to users exploiting STAR-RIS-enabled transmission and reflection links. Considering both unicast and multicast communication, we formulate a joint active and passive beamforming optimization problem for each of the proposed operating protocols for minimization of the power consumption of the BS, while satisfying the quality-of-service (QoS) requirements of the users.
  \item For ES, we first transform the resulting highly-coupled non-convex optimization problem into a tractable form with decoupled optimization variables. To solve the transformed non-convex problem, we develop an efficient iterative algorithm by exploiting the penalty method and successive convex approximation (SCA). The algorithm is guaranteed to converge to a stationary point of the original optimization problem. Furthermore, we extend the proposed penalty-based iterative algorithm to solve the mixed-integer non-convex optimization problem obtained for MS. For TS, we show that the optimization problem can be decomposed into two subproblems, which can be efficiently solved using state-of-the-art algorithms and convex optimization techniques.
  \item Our numerical results unveil that 1) TS is generally preferable for unicast communication, while ES is superior for multicast communication; 2) the performance gain of STAR-RISs over conventional RISs increases with the number of RIS elements; 3) for ES, the performance gain of element-based amplitude control over group-based amplitude control is significant for unicast communication, while it is negligible for multicast communication.
\end{itemize}
\vspace{-0.4cm}
\subsection{Organization and Notation}
\vspace{-0.2cm}
The rest of this paper is organized as follows: Section II introduces the basic signal model for STAR-RISs, based on which three practical protocols for operating STAR-RISs are proposed. Section III presents the system model and the joint beamforming optimization problem formulations for STAR-RIS aided communication systems for both unicast and multicast transmission. Efficient algorithms are developed for solving the problems formulated for each proposed operating protocol in Section IV. Section V provides numerical results to verify the effectiveness of the proposed STAR-RIS designs compared to baseline schemes. Finally, Section VI concludes the paper.\\
\indent \emph{Notations:} Scalars, vectors, and matrices are denoted by lower-case, bold-face lower-case, and bold-face upper-case letters, respectively. ${\mathbb{C}^{N \times 1}}$ denotes the space of $N \times 1$ complex-valued vectors. ${{\mathbf{a}}^H}$ and $\left\| {\mathbf{a}} \right\|$ denote the conjugate transpose and the Euclidean norm of vector ${\mathbf{a}}$, respectively. ${\rm {diag}}\left( \mathbf{a} \right)$ denotes a diagonal matrix with the elements of vector ${\mathbf{a}}$ on the main diagonal. The distribution of a circularly symmetric complex Gaussian (CSCG) random variable with mean $\mu $ and variance ${\sigma ^2}$ is denoted by ${\mathcal{CN}}\left( {\mu,\sigma ^2} \right)$. ${{\mathbf{1}}_{m \times n}}$ and ${{\mathbf{0}}_{m \times n}}$ denote the all-one and all-zero matrices of size ${m \times n}$, respectively. ${\mathbb{H}^{N}}$ denotes the set of all $N$-dimensional complex Hermitian matrices. ${\rm {Rank}}\left( \mathbf{A} \right)$ and ${\rm {Tr}}\left( \mathbf{A} \right)$ denote the rank and the trace of matrix $\mathbf{A}$, respectively. ${\rm {Diag}}\left( \mathbf{A} \right)$ denotes a vector whose elements are extracted from the main diagonal elements of matrix $\mathbf{A}$. ${{\mathbf{A}}} \succeq 0$ indicates that $\mathbf{A}$ is a positive semidefinite matrix. ${\left\| {\mathbf{A}} \right\|_*}$, ${\left\| {\mathbf{A}} \right\|_2}$, and ${\left\| {\mathbf{A}} \right\|_F}$ denote the nuclear norm, spectral norm, and Frobenius norm of matrix $\mathbf{A}$, respectively.
\vspace{-0.2cm}
\section{Basic Signal Model and Practical Operating Protocols For STAR-RISs}
In this section, we present the basic signal model, and three practical operating protocols for STAR-RISs in wireless communication systems.
\vspace{-0.6cm}
\subsection{Basic Signal Model of STAR-RISs}
\vspace{-0.2cm}
As shown in Fig. \ref{STR}, the wireless signal incident on a given element of the STAR-RIS is divided into a transmitted and a reflected signal. To characterize this STAR feature, let $s_m$ denote the signal incident on the $m$th element of the STAR-RIS, where $m \in {\mathcal{M}} \triangleq \left\{ {1,2, \ldots ,M} \right\}$ and $M$ denotes the total number of elements. The signals transmitted and reflected by the $m$th element can be modelled as ${t_m} = \left( {\sqrt {\beta _m^t} {e^{j\theta _m^t}}} \right){s_m}$ and ${r_m} = \left( {\sqrt {\beta _m^r} {e^{j\theta _m^r}}} \right){s_m}$~\cite{STAR}, respectively, where $\sqrt {\beta _m^t}  \in \left[ {0,1} \right],\theta _m^t \in \left[ {0,2\pi } \right)$ and $\sqrt {\beta _m^r}  \in \left[ {0,1} \right],\theta _m^r \in \left[ {0,2\pi } \right)$ denote the amplitude and phase shift response\footnote{In this paper, the amplitude and phase shift coefficients are assumed to be continuously adjustable to be able to determine the maximum performance. For practical hardware implementations, the obtained continuous solutions can be quantized to discrete values. It has shown that the performance degradation caused by phase shift quantization is small when the resolution is larger than 3 bits~\cite{Mu_IRS}.} of the $m$th element's transmission and reflection coefficients\footnote{To investigate the maximum performance gain of STAR-RISs, we assume that the amplitudes and phase shifts of the transmission and reflection coefficients can be adjusted independently. However, in practical implementations, the amplitude and phase shift adjustment may be coupled, thus leading to a performance loss. Investigating this loss is beyond the scope of this work.}. Note that, for each element, the phase shifts for transmission and reflection (i.e., $\theta _m^t$ and $\theta _m^r$) can be generally chosen independent from each other~\cite{STAR}. However, the amplitude adjustments for transmission and reflection (i.e., $\sqrt {\beta _m^t} $ and $\sqrt {\beta _m^r} $) are coupled by the law of energy conservation. This means that, for each element, the sum of the energies of the transmitted and reflected signals has to be equal to the energy of the incident signal, i.e., ${\left| {{t_m}} \right|^2} + {\left| {{r_m}} \right|^2} = {\left| {{s_m}} \right|^2}$, which leads to the following condition for the transmission and reflection amplitude coefficients of each element\footnote{In this paper, we assume that the STAR-RIS does not impose a power loss. Our proposed solutions are also applicable to the case with power loss, i.e., $\beta _m^t + \beta _m^r = c,\forall m \in {\mathcal{M}}$, where $0<c<1$.}~\cite{STAR}:
\vspace{-0.4cm}
\begin{align}\label{amplitude condition}
\beta _m^t + \beta _m^r = 1,\forall m \in {\mathcal{M}}.
\end{align}
\vspace{-1.8cm}
\subsection{Three Practical Protocols for Operating STAR-RISs}
\vspace{-0.2cm}
As can be observed from \eqref{amplitude condition}, by properly adjusting the amplitude coefficients for transmission and reflection, a given element of a STAR-RIS can be operated in the full transmission mode (i.e., $\beta _m^t = 1,\beta _m^r = 0$, referred to as T mode), the full reflection mode (i.e., $\beta _m^t = 0,\beta _m^r = 1$, referred to as R mode), and the general simultaneous transmission and reflection mode ($\beta _m^t,\beta _m^r \in \left[ {0,1} \right]$, referred to as T\&R mode). Inspired by these observations, in this subsection, we propose three practical protocols for operating STAR-RISs in wireless communication systems, namely energy splitting (ES), mode switching (MS), and time switching (TS), as illustrated in Fig. \ref{structure}.
\subsubsection{Energy Splitting} For ES, as shown in Fig. \ref{ES}, all elements of the STAR-RIS are assumed to operate in the T\&R mode, where the energy of the signal incident on each element is generally split into the energies of the transmitted and reflected signals with an energy splitting ratio of $\beta _m^t:\beta _m^r$. In this case, the transmission- and reflection-coefficient matrices of the STAR-RIS are given by ${\mathbf{\Theta }}_t^{{\rm{ES}}} = {\rm{diag}}\left( {\sqrt {\beta _1^t} {e^{j\theta _1^t}},\sqrt {\beta _2^t} {e^{j\theta _2^t}}, \ldots ,\sqrt {\beta _M^t} {e^{j\theta _M^t}}} \right)$ and ${\mathbf{\Theta }}_r^{{\rm{ES}}} = {\rm{diag}}\left( {\sqrt {\beta _1^r} {e^{j\theta _1^r}},\sqrt {\beta _2^r} {e^{j\theta _2^r}}, \ldots ,\sqrt {\beta _M^r} {e^{j\theta _M^r}}} \right)$, respectively, where $\beta _m^t,\beta _m^r \in \left[ {0,1} \right]$, $\beta _m^t + \beta _m^r = 1$, and $\theta _m^t,\theta _m^r \in \left[ {0,2\pi } \right),\forall m \in {\mathcal{M}}$. For ES, since both the transmission and reflection coefficients of each element can be optimized, a high degree of flexibility for communication system design is enabled. However, the large number of design variables also cause a relatively high overhead for configuration information exchange between the BS and the STAR-RIS.
\subsubsection{Mode Switching} 
\begin{figure}[t!]
\centering
\subfigure[Energy splitting (ES).]{\label{ES}
\includegraphics[width= 2in]{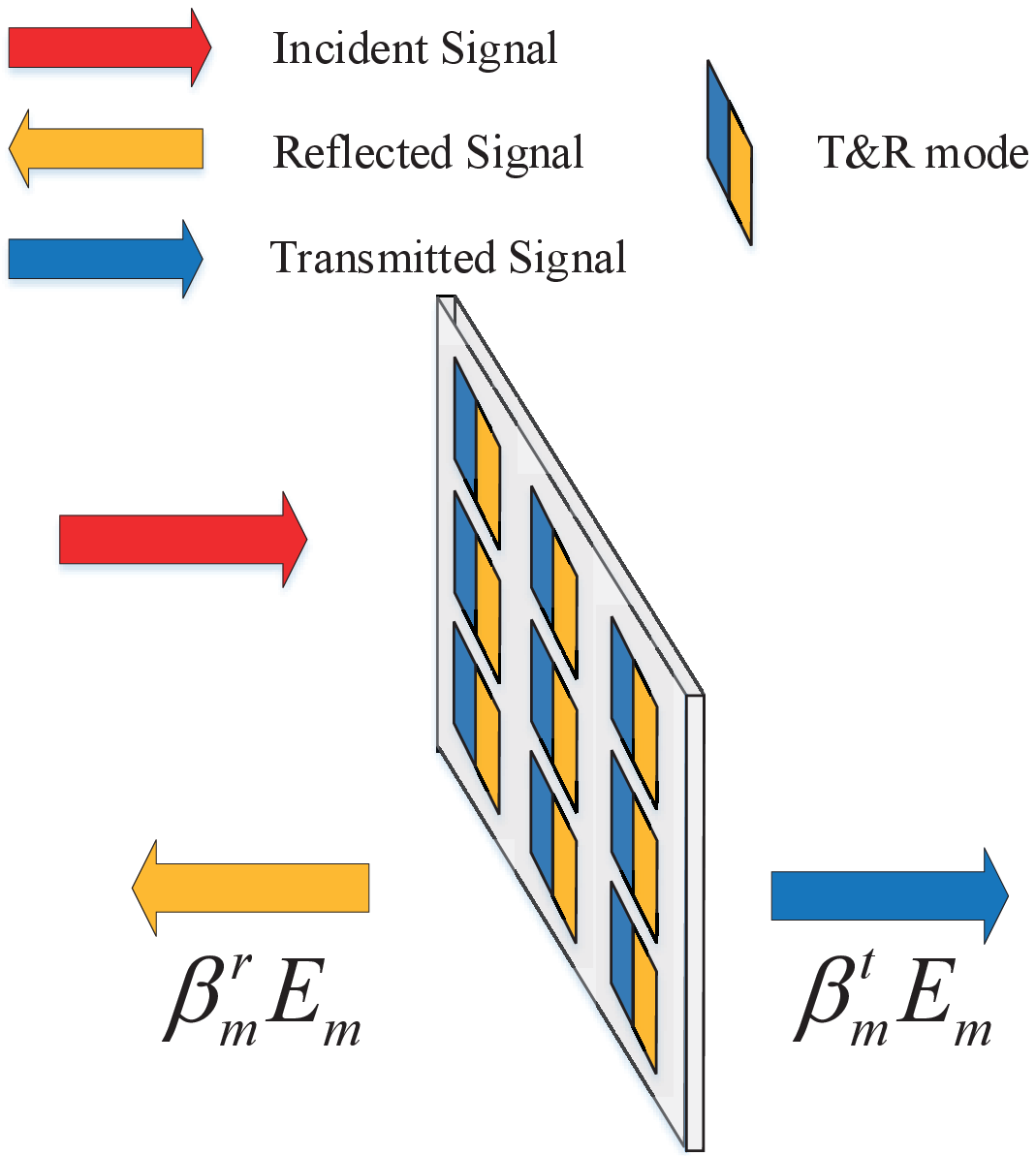}}
\subfigure[Mode switching (MS).]{\label{MS}
\includegraphics[width= 2in]{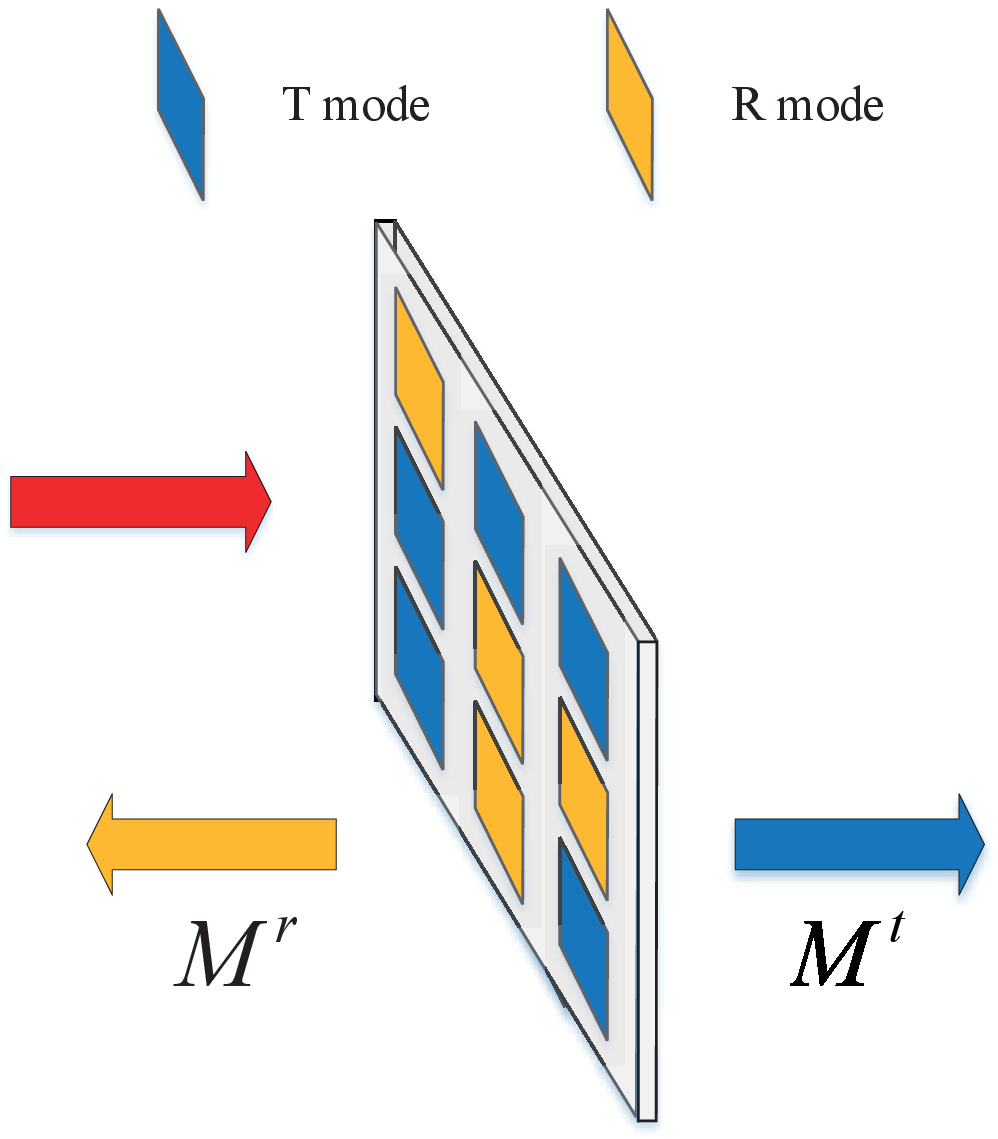}}
\subfigure[Time switching (TS).]{\label{TS}
\includegraphics[width= 2in]{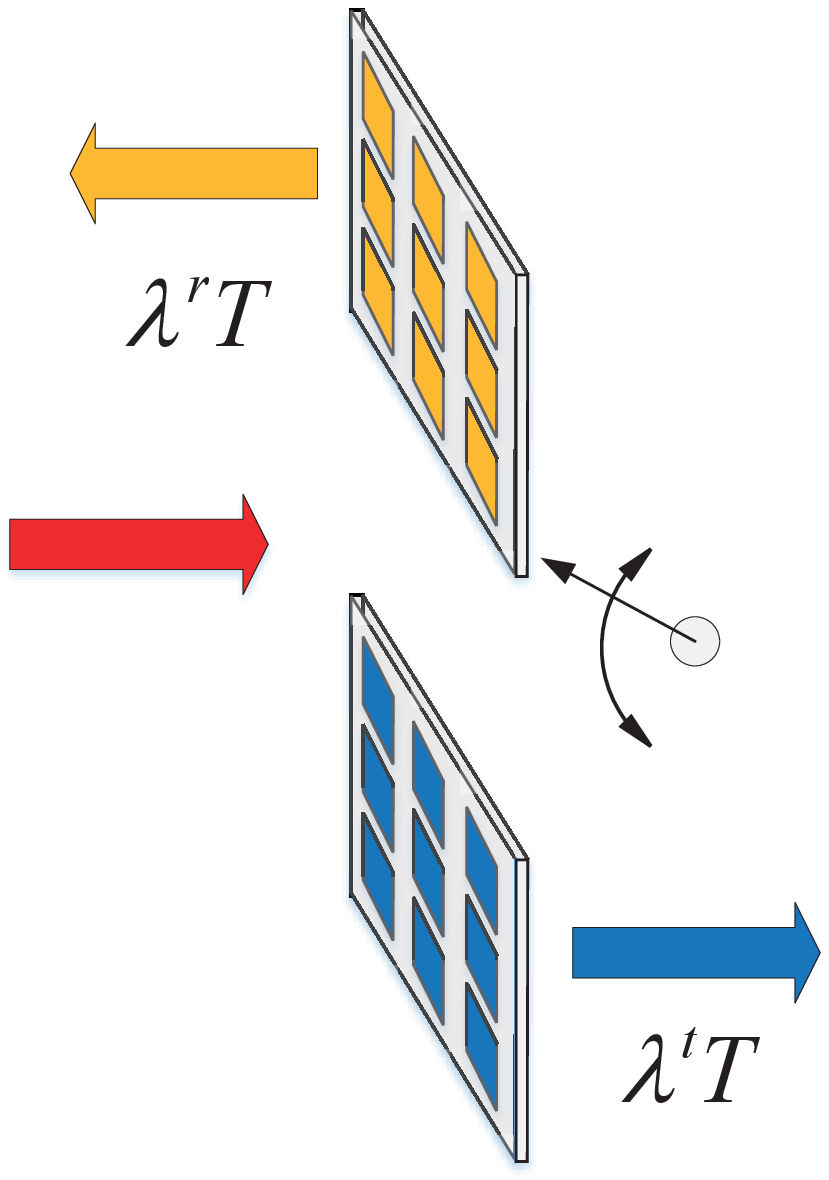}}
\caption{Three practical protocols for operating STAR-RISs, where $E_m$ denotes the energy of the signal incident on the $m$th element and $T$ denotes the total communication duration.}\label{structure}
\end{figure}

For MS, as shown in Fig. \ref{MS}, all elements of the STAR-RIS are divided into two groups. Specifically, one group contains $M^t$ elements that operate in the T mode, while the other group contains $M^r$ elements operating in the R mode, where ${M^t} + {M^r} = M$. Accordingly, the STAR-RIS transmission- and reflection-coefficient matrices are given by ${\mathbf{\Theta }}_t^{{\rm{MS}}} \!\!=\! {\rm{diag}}\!\left(\! {\sqrt {\beta _1^t} {e^{j\theta _1^t}},\!\sqrt {\beta _2^t} {e^{j\theta _2^t}},\! \ldots ,\!\sqrt {\beta _M^t} {e^{j\theta _M^t}}} \!\right)$ and ${\mathbf{\Theta }}_r^{{\rm{MS}}} \!\!=\! {\rm{diag}}\!\left(\! {\sqrt {\beta _1^r} {e^{j\theta _1^r}},\!\sqrt {\beta _2^r} {e^{j\theta _2^r}},\! \ldots ,\!\sqrt {\beta _M^r} {e^{j\theta _M^r}}} \!\right)$, respectively, where $\beta _m^t,\beta _m^r \in \left\{ {0,1} \right\}$, $\beta _m^t + \beta _m^r = 1$, and $\theta _m^t,\theta _m^r \in \left[ {0,2\pi } \right),\forall m \in {\mathcal{M}}$. We note that MS STAR-RISs can be regarded as a special case of ES STAR-RISs, where the amplitude coefficients for transmission and reflection are restricted to binary values. Therefore, MS generally cannot achieve the same full-dimension transmission and reflection beamforming gain as ES, since only a subset of the elements are selected for transmission and reflection, respectively. Nevertheless, MS is still appealing in practice, since such an ``on-off'' type operating protocol is much easier to implement compared to the ES protocol.
\subsubsection{Time Switching} Different from ES and MS, the TS STAR-RIS exploits the time domain and periodically switches all elements between the T mode and the R mode in different orthogonal time slots (referred to as T and R period), as illustrated in Fig. \ref{TS}. Let $0 \le{\lambda ^t}\le 1$ and $0 \le{\lambda ^r}\le 1$ denote the percentage of communication time allocated to the T period and R period, respectively, where ${\lambda ^t} + {\lambda ^r} = 1$. Therefore, the corresponding STAR-RIS transmission- and reflection-coefficient matrices are given by ${\mathbf{\Theta }}_{{t}}^{{\rm{TS}}} = {\rm{diag}}\left( {{e^{j\theta _1^t}},{e^{j\theta _2^t}}, \ldots ,{e^{j\theta _M^t}}} \right)$ and ${\mathbf{\Theta }}_{{r}}^{{\rm{TS}}} = {\rm{diag}}\left( {{e^{j\theta _1^r}},{e^{j\theta _2^r}}, \ldots ,{e^{j\theta _M^r}}} \right)$, where $\theta _m^t,\theta _m^r \in \left[ {0,2\pi } \right),\forall m \in {\mathcal{M}}$. Different from the ES and MS protocols, due to the exploitation of the time domain, the design of the transmission and reflection coefficients for TS is not coupled, and thus, easier to handle. However, the periodical switching of the elements introduces stringent requirements for time synchronization, which entails a higher hardware implementation complexity.\\
\indent In Table \ref{table:structure}, we summarize the optimization variables and constraints for the ES, MS, and TS operating protocols.
\begin{table*}[htbp]\scriptsize
\caption{Summary of optimization variables and constraints for the considered operating protocol.}
\vspace{-0.4cm}
\begin{center}
\centering
\resizebox{\textwidth}{!}{
\begin{tabular}{|l|l|l|l|}
\hline
\centering
\makecell[c]{\textbf{Optimization Variables}}  & \makecell[c]{\textbf{ES}} &\makecell[c]{\textbf{MS}} & \makecell[c]{\textbf{TS}} \\
\hline
\centering
\makecell[c]{Phase-shift coefficients}& \multicolumn{3}{c|}{\makecell[c]{$\theta _m^t,\theta _m^r \in \left[ {0,2\pi } \right)$}}    \\
\hline
\centering
\makecell[c]{Amplitude coefficients}&  \makecell[c]{$\beta _m^t,\beta _m^r \in \left[ {0,1} \right]$\\ $\beta _m^t + \beta _m^r = 1$}  &  \makecell[c]{$\beta _m^t,\beta _m^r \in \left\{ {0,1} \right\}$\\ $\beta _m^t + \beta _m^r = 1$} & \diagbox{\qquad \qquad \qquad}{\qquad \qquad \qquad} \\
\hline
\centering
\makecell[c]{Time allocation}&  \diagbox{\qquad \qquad \qquad}{\qquad \qquad \qquad}  & \diagbox{\qquad \qquad \qquad}{\qquad \qquad \qquad}   & \makecell[c]{$\lambda ^t,\lambda ^r \in \left[ {0,1} \right]$\\${\lambda ^t} + {\lambda ^r} = 1$}  \\
\hline
\end{tabular}
}
\end{center}
\label{table:structure}
\end{table*}%105
\vspace{-0.6cm}
\section{System Model and Problem Formulation}
Based on the proposed operating protocols, in this section, we present the system model of a STAR-RIS aided downlink communication system and formulate joint active and passive beamforming optimization problems for both unicast and multicast transmission.
\vspace{-0.4cm}
\subsection{System Model}
\vspace{-0.2cm}
\begin{figure}[t!]
  \centering
  \includegraphics[width=3in]{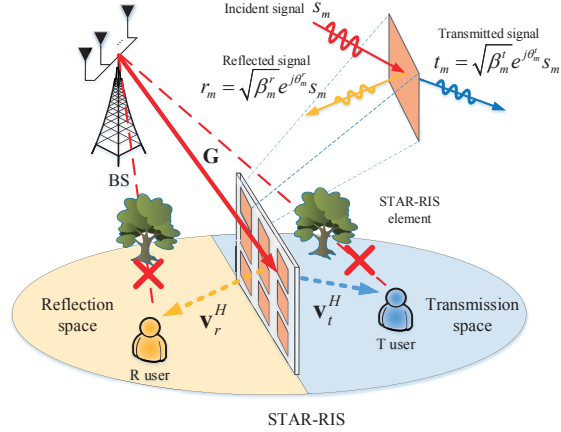}\\
  \caption{Illustration of a STAR-RIS aided downlink communication system, where the direct BS-user links are blocked by obstacles.}\label{model}
\end{figure}
As shown in Fig. \ref{model}, we consider a narrow-band STAR-RIS aided downlink communication system operating over frequency-flat channels, where an $N$-antenna BS communicates with multiple single-antenna users with the aid of a STAR-RIS comprising $M$ STAR elements. As previously discussed, the STAR-RIS can create full-space coverage by simultaneously transmitting and reflecting the incident signal. We refer to users that are located in the transmission space as T users, while users that are located in the reflection space are referred to as R users. As depicted in Fig. \ref{model}, in this paper, we assume that the direct communication links between the BS and the users are blocked by obstacles\footnote{The algorithms proposed in this work are also applicable for the case where the direct links between the BS and the T and R users exist.}, as this is one of the most challenging scenarios for conventional communication systems. The STAR-RIS is deployed to provide communication service for the users in the resulting signal dead zone by establishing additional transmission and reflection links. For simplicity of presentation and revealing fundamental design insights, in the remainder of the paper, we consider a two-user setup\footnote{The proposed algorithms can be extended to systems with multiple T and R users, see \textbf{Remark \ref{multiuser ES MS}} and \textbf{Remark \ref{multiuser TS}} in Section IV.}, i.e., the STAR-RIS aided communication system serves one T user and one R user. Let ${\mathbf{G}} \in {{\mathbb{C}}^{M \times N}}$, ${\mathbf{v}}_t^H \in {{\mathbb{C}}^{1 \times M}}$, and ${\mathbf{v}}_r^H \in {{\mathbb{C}}^{1 \times M}}$ represent the narrow-band quasi-static fading channels from the BS to the STAR-RIS, the STAR-RIS to T user, and the STAR-RIS to R user, respectively. In order to be able to unveil the maximum performance gain enabled by STAR-RISs, the perfect CSI of all channels is assumed to be available at the BS\footnote{Because of the nearly-passive mode of operation of RISs, channel estimation in RIS-assisted wireless systems is a challenging task. To address this issue, numerous efficient channel estimation schemes have been proposed for conventional RISs~\cite{He_CSI,Deepak_CSI,Hang_CSI,Wang_CSI,You,Zheng_OFDMA}, which can also be readily employed for STAR-RISs. For instance, using the TS protocol, the CSI of the T and R users can be consecutively estimated using existing channel acquisition methods~\cite{He_CSI,Deepak_CSI,Hang_CSI,Wang_CSI,You,Zheng_OFDMA}. However, the development of more efficient CSI estimation techniques for STAR-RISs based on the ES and MS protocols to simultaneously acquire the CSI of the T and R users is an interesting topic for future work.}.\\
\indent In this paper, we study the joint design of the active beamforming at the BS and the passive transmission and reflection beamforming at the STAR-RIS for both unicast and multicast communication. For unicast, the BS sends independent information to the T and R users. For multicast, the BS sends the same information to both users. For both scenarios, the three proposed STAR-RIS operating protocols are considered.
\vspace{-0.6cm}
\subsection{Unicast Communication and Problem Formulation}
\vspace{-0.2cm}
In unicast transmission, let ${{\mathbf{w}}_k}$ and ${{{x}}_k}$ denote the active beamforming vector and the information-bearing symbol for user $k \in \left\{ {t,r} \right\}$ at the BS, respectively.
\subsubsection{ES and MS} When the ES or MS protocols are employed at the STAR-RIS, the received signal at user $k\in \left\{ {t,r} \right\}$ is given by
\vspace{-0.4cm}
\begin{align}\label{received signal}
y_{{\rm{UC}},k}^{\rm{ES/MS}} = {\mathbf{v}}_k^H{\mathbf{\Theta }}_k^{\rm{ES/MS}}{\mathbf{G}}\left( {{{\mathbf{w}}_t}{x_t} + {{\mathbf{w}}_r}{x_r}} \right) + {n_k},
\end{align}
\vspace{-1.2cm}

\noindent where ${\mathbb{E}}\left[ {{{\left| {{x_k}} \right|}^2}} \right] = 1$ and ${n_k} \sim {\mathcal{CN}}\left( {0,\sigma _k^2} \right)$ denotes the additive white Gaussian noise (AWGN) at user $k$. Therefore, the achievable communication rate of user $k \in \left\{ {t,r} \right\}$ for ES and MS is given by
\vspace{-0.4cm}
\begin{align}\label{ES1 rate}
  R_{{\rm{UC}},k}^{\rm{ES/MS}} = {\log _2}\left( {1 + \frac{{{{\left| {{\mathbf{v}}_k^H{\mathbf{\Theta }}_k^{\rm{ES/MS}}{\mathbf{G}}{{\mathbf{w}}_k}} \right|}^2}}}{{{{\left| {{\mathbf{v}}_k^H{\mathbf{\Theta }}_k^{\rm{ES/MS}}{\mathbf{G}}{{\mathbf{w}}_{\overline k}}} \right|}^2} + \sigma _k^2}}} \right),
\end{align}
\vspace{-0.8cm}

\noindent where $\overline k  = r$, if $k = t$; and $\overline k  = t$, otherwise.
\subsubsection{TS} For TS, the BS consecutively sends information to the two users in the T and R periods. Thus, the corresponding achievable communication rate of user $k \in \left\{ {t,r} \right\}$ can be expressed as
\vspace{-0.4cm}
\begin{align}\label{TS rate}
R_{{\rm{UC}},k}^{\rm{TS}} = {\lambda ^k}{\log _2}\left( {1 + \frac{{{{\left| {{\mathbf{v}}_k^H{\mathbf{\Theta }}_{{k}}^{\rm{TS}}{\mathbf{G}}{{\mathbf{w}}_k}} \right|}^2}}}{{{\lambda ^k}\sigma _k^2}}} \right),
\end{align}
\vspace{-1cm}

\noindent where ${\lambda ^k}$ in \eqref{TS rate} is due to the fact that the BS sends information to user $k$ employing ${{\mathbf{w}}_k}$ in ${\lambda ^k}$ fraction of the total communication time. The transmit power of user $k$ is increased by ${1 \mathord{\left/
 {\vphantom {1 {{\lambda ^k}}}} \right.
 \kern-\nulldelimiterspace} {{\lambda ^k}}}$ to ensure a fair comparison with the ES and MS protocols.\\
\indent For unicast transmission, we aim to minimize the total power consumption of the BS by jointly optimizing the active beamforming at the BS and the passive transmission and reflection beamforming at the STAR-RIS for a given operating protocol, while satisfying the QoS requirement of both users. Then, the optimization problem can be formulated as follows:
\vspace{-0.3cm}
\begin{subequations}\label{P1 X}
\begin{align}
&\mathop {\min }\limits_{ {{{\mathbf{w}}_k},{\mathbf{\Theta }}_k^{\rm{X}},{\lambda ^k}} } \;\;\left\| {{{\mathbf{w}}_t}} \right\|^2 + \left\| {{{\mathbf{w}}_r}} \right\|^2  \\
\label{QoS X}{\rm{s.t.}}\;\;& R_{{\rm{UC}},k}^{\rm{X}}\ge {\overline R _k},\forall k \in \left\{ {t,r} \right\},\\
\label{Theta X}&{\mathbf{\Theta }}_k^{\rm{X}} \in {{\mathcal{F}}^{\rm{X}}},\forall k \in \left\{ {t,r} \right\},\\
\label{lambda TS0}&0 \le {\lambda ^t} \le 1,0 \le {\lambda ^r} \le 1,{\lambda ^t} + {\lambda ^r} = 1,
\end{align}
\end{subequations}
\vspace{-1.2cm}

\noindent where ${\rm{X}} \in \left\{ {{\rm{ES}},{\rm{MS}},{\rm{TS}}} \right\}$ indicates the employed STAR-RIS operating protocol, ${{{\mathcal{F}}}^{\rm{X}}}$ characterizes the corresponding feasible set for the transmission- and reflection-coefficient matrices, and ${\overline R _k}$ denotes the minimum rate requirement of user $k$. The time allocation variables, $\left\{ {{\lambda ^k}} \right\}$, and constraint \eqref{lambda TS0} are only valid when the TS protocol is employed, i.e., ${\rm{X}}={\rm{TS}}$. Let ${\mathbf{H}} = \left[ {{{\mathbf{h}}_t}\;{{\mathbf{h}}_r}} \right] \in {{\mathbb{C}}^{N \times 2}}$, where ${\mathbf{h}}_k^H = {\mathbf{v}}_k^H{\mathbf{\Theta }}_k^{{\rm{ES/MS}}}{\mathbf{G}},\forall k \in \left\{ {t,r} \right\}$. For the ES and MS protocols, assuming that $N\ge2$, a sufficient condition for the feasibility of problem \eqref{P1 X} for any finite ${\overline R _k}$ is that ${\rm{Rank}}\left( {\mathbf{H}} \right) = 2$, where a zero-forcing solution can be easily constructed for the active beamforming~\cite{Wu2019IRS}. Given the independent STAR-RIS-user channels, $\left\{ {{\mathbf{v}}_k^H} \right\}$, and the different transmission- and reflection-coefficient matrices, $\left\{ {{\mathbf{\Theta }}_k^{{\rm{ES/MS}}}} \right\}$, such a rank condition can be satisfied with a high probability. For the TS operating protocol, problem \eqref{P1 X} is always feasible for any finite ${\overline R _k}$ since there is no inter-user interference during the T and R periods, respectively.
\vspace{-0.3cm}
\begin{remark}\label{difference}
\emph{Compared to the power consumption minimization problem for conventional reflecting-only RISs~\cite{Wu2019IRS}, the main challenges for solving \eqref{P1 X} can be summarized as follows. First, STAR-RISs require the optimization of two types of passive beamforming (i.e., transmission and reflection beamforming), while for conventional reflecting-only RISs, only the reflection beamforming has to be designed. Mathematically, the optimization problem for conventional reflecting-only RISs is a special case of optimization problem \eqref{P1 X} for STAR-RISs, where the transmission function is turned off and only the R user is served. Second, for the proposed ES and MS protocols, transmission and reflection beamforming is coupled together, which further complicates resource allocation compared to conventional reflecting-only RISs. Therefore, the considered problem \eqref{P1 X} for STAR-RISs is much more challenging to solve than that for conventional reflecting-only RISs.}
\end{remark}
\vspace{-1cm}
\subsection{Multicast Communication and Problem Formulation}
\vspace{-0.2cm}
\subsubsection{ES and MS} For ES and MS, in multicast transmission, the BS employs one active beamforming vector, denoted by ${{\mathbf{w}}_c}$, to convey the same symbol, ${{{s}}_c}$, to the T and R users. As a result, the corresponding received signal at user $k \in \left\{ {t,r} \right\}$ is given by
\vspace{-0.3cm}
\begin{align}\label{received signal mc}
y_{{\rm{MC}},k}^{\rm{ES/MS}} = {\mathbf{v}}_k^H{\mathbf{\Theta }}_k^{\rm{ES/MS}}{\mathbf{G}} {{{\mathbf{w}}_c}{s_c}} + {n_k}.
\end{align}
\vspace{-1.2cm}

\noindent Accordingly, the achievable communication rate of user $k \in \left\{ {t,r} \right\}$ can be expressed as
\vspace{-0.3cm}
\begin{align}\label{ES1 rate mc}
R_{{\rm{MC}},k}^{\rm{ES/MS}} = {\log _2}\left( {1 + \frac{{{{\left| {{\mathbf{v}}_k^H{\mathbf{\Theta }}_k^{{\rm{ES/MS}}}{\mathbf{G}}{{\mathbf{w}}_c}} \right|}^2}}}{{\sigma _k^2}}} \right).
\end{align}
\vspace{-0.8cm}
\subsubsection{TS} Since for TS the users are served in different orthogonal time slots, the BS can still employ different active beamforming vectors, which are denoted by ${{\mathbf{w}}_{c,t}}$ and ${{\mathbf{w}}_{c,r}}$, to send the same symbol to the T and R users. Therefore, for multicast communication with TS, the achievable communication rate at user $k \in \left\{ {t,r} \right\}$ is given by
\vspace{-0.3cm}
\begin{align}\label{TS rate MC}
R_{{\rm{MC}},k}^{\rm{TS}} = {\lambda ^k}{\log _2}\left( {1 + \frac{{{{\left| {{\mathbf{v}}_k^H{\mathbf{\Theta }}_{{k}}^{\rm{TS}}{\mathbf{G}}{{\mathbf{w}}_{c,k}}} \right|}^2}}}{{{\lambda ^k}\sigma _k^2}}} \right).
\end{align}
\vspace{-1cm}

\noindent We note that, for TS, the multicast communication rate in \eqref{TS rate MC} is identical to the unicast communication rate in \eqref{TS rate}. This is because, regardless of whether unicast or multicast communication is considered, the proposed TS protocol can serve only one user in each time instant. \\ %for all employed operating protocols,
\indent Since the information has to be delivered to both users, the system performance of multicast communication is limited by the user achieving the smaller communication rate. As a result, the effective system multicasting rate for the considered operating protocols is given by
\vspace{-0.4cm}
\begin{align}\label{MC system rate}
R_{{\rm{MC}}}^{\rm{X}} = \min \left\{ {R_{{\rm{MC}},t}^{\rm{X}},R_{{\rm{MC}},r}^{\rm{X}}} \right\},
\end{align}
\vspace{-1.2cm}

\noindent where ${\rm{X}} \in \left\{ {{\rm{ES}},{\rm{MS}},{\rm{TS}}} \right\}$. Then, for multicast transmission, the joint beamforming optimization problem for ES, MS, and TS can be formulated as
\vspace{-0.3cm}
\begin{subequations}\label{P2 X}
\begin{align}
&{{\mathop {\min }\limits_{{{\mathbf{w}}_c}, {{\mathbf{\Theta }}_k^{{\rm{ES/MS}}}} } \;\;{{\left\| {{{\mathbf{w}}_c}} \right\|}^2}} \mathord{\left/
 {\vphantom {{\mathop {\min }\limits_{{{\mathbf{w}}_c}, {{\mathbf{\Theta }}_k^{{\rm{ES/MS}}}} } \;\;{{\left\| {{{\mathbf{w}}_c}} \right\|}^2}} {\mathop {\min }\limits_{ {{{\mathbf{w}}_{c,k}},{\mathbf{\Theta }}_k^{\rm{TS}},{\lambda ^k}} } \;\;{{\left\| {{{\mathbf{w}}_{c,t}}} \right\|}^2} + {{\left\| {{{\mathbf{w}}_{c,r}}} \right\|}^2}}}} \right.
 \kern-\nulldelimiterspace} {\mathop {\min }\limits_{ {{{\mathbf{w}}_{c,k}},{\mathbf{\Theta }}_k^{\rm{TS}},{\lambda ^k}} } \;\;{{\left\| {{{\mathbf{w}}_{c,t}}} \right\|}^2} + {{\left\| {{{\mathbf{w}}_{c,r}}} \right\|}^2}}} \\
\label{QoS 2X}{\rm{s.t.}}\;\;& R_{{\rm{MC}}}^{\rm{X}}\ge {\overline R _c},\\
\label{Theta 2X}&{\mathbf{\Theta }}_k^{\rm{X}} \in {{\mathcal{F}}^{\rm{X}}},\\
\label{lambda TS00}&0 \le {\lambda ^t} \le 1,0 \le {\lambda ^r} \le 1,{\lambda ^t} + {\lambda ^r} = 1,
\end{align}
\end{subequations}
\vspace{-1.2cm}

\noindent where \eqref{QoS 2X} denotes the minimum system multicast rate constraint with required communication rate ${\overline R _c}$. We note that $\left\{ {{\lambda ^k}} \right\}$ and constraint \eqref{lambda TS00} are only valid for TS. As there is no inter-user interference in multicast, for each operating protocol, problem \eqref{P2 X} is always feasible for any finite ${\overline R _c}$.
\vspace{-0.7cm}
\subsection{Discussion}
\vspace{-0.2cm}
Note that the formulated joint beamforming optimization problems for multicast transmission can be regarded as special cases of those for unicast transmission. More specifically, for ES and MS, problem \eqref{P2 X} for multicast transmission can be obtained from problem \eqref{P1 X} for unicast transmission by setting ${{\mathbf{w}}_t} = {{\mathbf{w}}_r} = {{\mathbf{w}}_c}$, ${\overline R _t}={\overline R _r}={\overline R _c}$ and removing the inter-user interference term in the denominator of the unicast communication rate expression in \eqref{ES1 rate}. For TS, problem \eqref{P2 X} for multicast transmission can be directly obtained from problem \eqref{P1 X} for unicast transmission by setting ${{\mathbf{w}}_t} ={{\mathbf{w}}_{c,t}}$, ${{\mathbf{w}}_r} ={{\mathbf{w}}_{c,r}}$, and ${\overline R _t}={\overline R _r}={\overline R _c}$. Motivated by this, in the following, we mainly focus on joint beamforming optimization problem \eqref{P1 X} for unicast transmission since problem \eqref{P2 X} for multicast transmission can be solved in a similar manner. However, problem \eqref{P1 X} is a non-convex optimization problem since the left-hand-side (LHS) of \eqref{QoS X} is not concave, as $\left\{ {{{\mathbf{\Theta }}_k^{\rm{X}}}} \right\}$ and $\left\{ {{{\mathbf{w}}_k}} \right\}$ are highly-coupled. Moreover, the feasible set, ${{\mathcal{F}}^{\rm{X}}}$, is in general non-convex. Therefore, it is difficult to obtain a globally optimal solution for such a challenging problem. In the following, we develop efficient algorithms to find a high-quality suboptimal solution for the three considered operating protocols.
\vspace{-0.2cm}
\section{Solution of Joint Beamforming Optimization Problems}
In this section, we first propose a penalty-based iterative algorithm for solving the joint beamforming optimization problem for ES. This algorithm is then further extended to solve the problem for MS. Then, we show that the problem for TS can be decomposed into two subproblems, namely a transmission/reflection coefficient design problem and a resource allocation problem, which can be solved in a relatively straightforward manner.
\vspace{-0.5cm}
\subsection{Proposed Solution for ES and MS}
\vspace{-0.2cm}
To start with, we first transform problem \eqref{P1 X} for ES and MS into a more tractable form. To facilitate the design, we define the transmission- and reflection-coefficient vectors for ES and MS as ${{\mathbf{q}}_k^{\rm{ES/MS}}} = {\left[ {\sqrt {\beta _1^k} {e^{j\theta _1^k}},\sqrt {\beta _2^k} {e^{j\theta _2^k}}, \ldots ,\sqrt {\beta _M^k} {e^{j\theta _M^k}}} \right]^H}, \forall k \in \left\{ {t,r} \right\}$, which leads to ${\left| {{\mathbf{v}}_k^H{\mathbf{\Theta }}_k^{\rm{ES/MS}}{\mathbf{G}}{{\mathbf{w}}_k}} \right|^2}\! =\! {\left| {{{\left( {{\mathbf{q}}_k^{\rm{ES/MS}}} \right)}^H}{{\mathbf{H}}_k}{{\mathbf{w}}_k}} \right|^2}$, where ${{\mathbf{H}}_k} \!= \!{\rm{diag}}\left( {{\mathbf{v}}_k^H} \right){\mathbf{G}}$. Moreover, we define ${\mathbf{Q}}_k^{\rm{ES/MS}} \!=\! {\mathbf{q}}_k^{\rm{ES/MS}}{\left(\! {{\mathbf{q}}_k^{\rm{ES/MS}}} \!\right)^H}\!,\!\forall k \!\in \!\left\{ {t,r} \right\}$, which satisfies ${{\mathbf{Q}}_k^{\rm{ES/MS}}}\!\succeq\! 0$, ${\rm{Rank}}\!\left(\! {{{\mathbf{Q}}_k^{\rm{ES/MS}}}} \!\right) \!= \!1$, and ${\rm{Diag}}\!\left(\! {{{\mathbf{Q}}_k^{\rm{ES/MS}}}} \!\right)\! = {{\bm{\beta}} ^k}$, where ${{\bm{\beta}} ^k} \triangleq \left[ {\beta _1^k,\beta _2^k, \ldots ,\beta _M^k} \right]$. Similarly, we define ${{\mathbf{W}}_k} = {{\mathbf{w}}_k}{\mathbf{w}}_k^H,\forall k \in \left\{ {t,r} \right\}$, which satisfies ${{\mathbf{W}}_k}\succeq 0$ and ${\rm{Rank}}\left( {{{\mathbf{W}}_k}} \right) = 1$. Then, optimization problem \eqref{P1 X} for ES and MS can be reformulated as follows:
\vspace{-0.6cm}
\begin{subequations}\label{P1 GES}
\begin{align}
&\mathop {\min }\limits_{ {{{\mathbf{W}}_k},{{\mathbf{Q}}_k^{\rm{Y}}},{{\bmm{\beta}} _k}} } \;\;{\rm{Tr}}\left( {{{\mathbf{W}}_t}} \right) + {\rm{Tr}}\left( {{{\mathbf{W}}_r}} \right)  \\
\label{QoS GES}{\rm{s.t.}}\;\;&{\overline \gamma  _k}{\rm{Tr}}\left( {{{\mathbf{Q}}_k^{\rm{Y}}}{{\mathbf{H}}_k}{{\mathbf{W}}_{\overline k }}{\mathbf{H}}_k^H} \right) - {\rm{Tr}}\left( {{{\mathbf{Q}}_k^{\rm{Y}}}{{\mathbf{H}}_k}{{\mathbf{W}}_{k}}{\mathbf{H}}_k^H} \right) + {\overline \gamma  _k}\sigma _k^2 \le 0,\forall k \in \left\{ {t,r} \right\},\\
\label{amplitude GES1}&{\rm{Diag}}\left( {{{\mathbf{Q}}_k^{\rm{Y}}}} \right) = {{\bm{\beta}} ^k},\forall k \in \left\{ {t,r} \right\},\\
\label{Rank Q GES}&{\rm{Rank}}\left( {{{\mathbf{Q}}_k^{\rm{Y}}}} \right) = 1,\forall k \in \left\{ {t,r} \right\},\\
\label{Rank W GES}&{\rm{Rank}}\left( {{{\mathbf{W}}_k}} \right) = 1,\forall k \in \left\{ {t,r} \right\},\\
\label{H Q W GES}&{{\mathbf{Q}}_k^{\rm{Y}}} \succeq 0,{{\mathbf{W}}_k} \succeq 0,\forall k \in \left\{ {t,r} \right\},\\
\label{amplitude GES2}&0 \le \beta _m^t,\beta _m^r \le 1,\beta _m^t + \beta _m^r = 1,\forall m \in {\mathcal{M}},\\
\label{AC}&\beta _m^k \in \left\{ {0,1} \right\},\forall k \in \left\{ {t,r} \right\},\forall m \in {\mathcal{M}},
\end{align}
\end{subequations}
\vspace{-1.2cm}

\noindent where ${\rm{Y}} \in \left\{ {{\rm{ES}},{\rm{MS}}} \right\}$ specifies the operating protocol and ${\overline \gamma _k} = {2^{{{\overline R}_k}}} - 1,\forall k \in \left\{ {t,r} \right\}$. We note that constraint \eqref{AC} is only present for MS, i.e., ${\rm{Y}} ={\rm{MS}}$. Problem \eqref{P1 GES} is still non-convex due to the highly-coupled non-convex constraint \eqref{QoS GES} and the non-convex rank-one constraint \eqref{Rank Q GES} and \eqref{Rank W GES}. Moreover, for MS, the optimization problem becomes a mixed-integer non-convex problem due to binary constraint \eqref{AC}. In the following, we first focus on optimization problem \eqref{P1 GES} for ES, when constraint \eqref{AC} is absent.
\subsubsection{Penalty-based Algorithm for ES}
\indent Before handling the highly-coupled non-convex constraint \eqref{QoS GES}, we first have the following lemma.
\vspace{-0.3cm}
\begin{lemma}\label{AB}
\emph{For any two Hermitian matrices ${\mathbf{A}} \in {{\mathbb{H}}^M}$ and ${\mathbf{B}} \in {{\mathbb{H}}^M}$ having the same size, we have the following two equalities:}
\vspace{-0.4cm}
\begin{subequations}
\begin{align}\label{D 1}
{\rm{Tr}}\left( {{\mathbf{AB}}} \right) &= \frac{1}{2}\left\| {{\mathbf{A}} + {\mathbf{B}}} \right\|_F^2 - \frac{1}{2}\left\| {\mathbf{A}} \right\|_F^2 - \frac{1}{2}\left\| {\mathbf{B}} \right\|_F^2,\\\label{D 2}
- {\rm{Tr}}\left( {{\mathbf{AB}}} \right) &= \frac{1}{2}\left\| {{\mathbf{A}} - {\mathbf{B}}} \right\|_F^2 - \frac{1}{2}\left\| {\mathbf{A}} \right\|_F^2 - \frac{1}{2}\left\| {\mathbf{B}} \right\|_F^2.
\end{align}
\end{subequations}
\begin{proof}
\vspace{-0.2cm}
\emph{We first prove equation \eqref{D 1}, whose right-hand-side (RHS) can be rewritten as}
\[\begin{gathered}
  \frac{1}{2}\left\| {{\mathbf{A}} + {\mathbf{B}}} \right\|_F^2 - \frac{1}{2}\left\| {\mathbf{A}} \right\|_F^2 - \frac{1}{2}\left\| {\mathbf{B}} \right\|_F^2 \hfill \\
   = \frac{1}{2}{\rm{Tr}}\left( {{{\left( {{\mathbf{A}} + {\mathbf{B}}} \right)}^H}\left( {{\mathbf{A}} + {\mathbf{B}}} \right)} \right) - \frac{1}{2}{\rm{Tr}}\left( {{{\mathbf{A}}^H}{\mathbf{A}}} \right) - \frac{1}{2}{\rm{Tr}}\left( {{{\mathbf{B}}^H}{\mathbf{B}}} \right) \hfill \\
   = \frac{1}{2}{\rm{Tr}}\left( {{{\mathbf{A}}^H}{\mathbf{A}}} \right) + \frac{1}{2}{\rm{Tr}}\left( {{{\mathbf{A}}^H}{\mathbf{B}}} \right) + \frac{1}{2}{\rm{Tr}}\left( {{{\mathbf{B}}^H}{\mathbf{A}}} \right) + \frac{1}{2}{\rm{Tr}}\left( {{{\mathbf{B}}^H}{\mathbf{B}}} \right) - \frac{1}{2}{\rm{Tr}}\left( {{{\mathbf{A}}^H}{\mathbf{A}}} \right) - \frac{1}{2}{\rm{Tr}}\left( {{{\mathbf{B}}^H}{\mathbf{B}}} \right) \hfill \\
   = \frac{1}{2}{\rm{Tr}}\left( {{{\mathbf{A}}^H}{\mathbf{B}}} \right) + \frac{1}{2}{\rm{Tr}}\left( {{{\mathbf{B}}^H}{\mathbf{A}}} \right) \mathop  = \limits^{\left( a \right)} {\rm{Tr}}\left( {{\mathbf{AB}}} \right), \hfill \\
\end{gathered} \]
\vspace{-0.6cm}

\noindent \emph{where $\left( a \right)$ is due to the fact that ${\mathbf{A}}$ and ${\mathbf{B}}$ are Hermitian matrices. Equation \eqref{D 2} can be proved in a similar manner, and we omit the details for brevity.}
\end{proof}
\end{lemma}
\vspace{-0.2cm}
Based on \textbf{Lemma \ref{AB}}, the first and second non-convex terms in \eqref{QoS GES} can be transformed into the following difference of convex (DC) functions:
\vspace{-0.5cm}
\begin{subequations}
\begin{align}\label{DC 1}
&{\overline \gamma  _k}{\rm{Tr}}\left( {{{\mathbf{Q}}_k^{\rm{ES}}}{{\mathbf{H}}_k}{{\mathbf{W}}_{\overline k}}{\mathbf{H}}_k^H} \right) = \frac{{{{\overline \gamma  }_k}}}{2}\left\| {{{\mathbf{Q}}_k^{\rm{ES}}} + {{\mathbf{H}}_k}{{\mathbf{W}}_{\overline k}}{\mathbf{H}}_k^H} \right\|_F^2 - \frac{{{{\overline \gamma  }_k}}}{2}\left\| {{{\mathbf{Q}}_k^{\rm{ES}}}} \right\|_F^2 - \frac{{{{\overline \gamma  }_k}}}{2}\left\| {{{\mathbf{H}}_k}{{\mathbf{W}}_{\overline k}}{\mathbf{H}}_k^H} \right\|_F^2 \triangleq {\Upsilon _{k\overline k}^{\rm{ES}}},\\
\label{DC 2}
&- {\rm{Tr}}\left( {{{\mathbf{Q}}_k^{\rm{ES}}}{{\mathbf{H}}_k}{{\mathbf{W}}_{k}}{\mathbf{H}}_k^H} \right) = \frac{1}{2}\left\| {{{\mathbf{Q}}_k^{\rm{ES}}} - {{\mathbf{H}}_k}{{\mathbf{W}}_{k}}{\mathbf{H}}_k^H} \right\|_F^2 - \frac{1}{2}\left\| {{{\mathbf{Q}}_k^{\rm{ES}}}} \right\|_F^2 - \frac{1}{2}\left\| {{{\mathbf{H}}_k}{{\mathbf{W}}_{k}}{\mathbf{H}}_k^H} \right\|_F^2 \triangleq {\Pi _{kk}^{\rm{ES}}}.
\end{align}
\end{subequations}
\vspace{-1.0cm}

\indent {Furthermore, the non-convex rank-one constraint \eqref{Rank Q GES} can be equivalently written as the following equality constraint~\cite{Tao_OTA}:
\vspace{-0.4cm}
\begin{align}\label{DC rank}
{\left\| {{{\mathbf{Q}}_k^{\rm{ES}}}} \right\|_*} - {\left\| {{{\mathbf{Q}}_k^{\rm{ES}}}} \right\|_2} = 0,\forall k \in \left\{ {t,r} \right\},
\end{align}}
\vspace{-1.2cm}

\noindent where ${\left\| {\mathbf{Q}}_k^{\rm{ES}} \right\|_ * } = \sum\nolimits_{i} {{\sigma _{i}}\left( {\mathbf{Q}}_k^{\rm{ES}} \right)}$ and ${\left\| {\mathbf{Q}}_k^{\rm{ES}} \right\|_2} = {\sigma _{1}}\left( {\mathbf{Q}}_k^{\rm{ES}} \right)$ denote the nuclear norm and the spectral norm, respectively, and ${\sigma _{i}}\left( {\mathbf{Q}}_k^{\rm{ES}} \right)$ is the $i$th largest singular value of matrix ${\mathbf{Q}}_k^{\rm{ES}}$. Note that for any ${{\mathbf{Q}}_k^{\rm{ES}}} \in {{\mathbb{H}}^M}$ and ${{\mathbf{Q}}_k^{\rm{ES}}} \succeq 0$, we always have ${\left\| {{{\mathbf{Q}}_k^{\rm{ES}}}} \right\|_*} - {\left\| {{{\mathbf{Q}}_k^{\rm{ES}}}} \right\|_2} \ge 0$, where equality holds if and only if ${{\mathbf{Q}}_k^{\rm{ES}}}$ is a rank-one matrix. Therefore, equality constraint \eqref{DC rank} is only met for rank-one matrices ${\left\{ {{{\mathbf{Q}}_k^{\rm{ES}}}} \right\}}$.\\
\indent Next, we employ the penalty method~\cite{penalty} to solve problem \eqref{P1 GES}. By exploiting \eqref{DC 1}, \eqref{DC 2}, and \eqref{DC rank}, we obtain the following optimization problem:
\vspace{-0.5cm}
\begin{subequations}\label{P1 GES1}
\begin{align}
\mathop {\min }\limits_{ {{{\mathbf{W}}_k},{{\mathbf{Q}}_k^{\rm{ES}}},{{\bmm{\beta}} _k}} } &{\rm{Tr}}\left( {{{\mathbf{W}}_t}} \right) + {\rm{Tr}}\left( {{{\mathbf{W}}_r}} \right) + \eta  {\sum\nolimits_{k \in \left\{ {t,r} \right\}} {\left( {{{\left\| {{{\mathbf{Q}}_k^{\rm{ES}}}} \right\|}_*} - {{\left\| {{{\mathbf{Q}}_k^{\rm{ES}}}} \right\|}_2}} \right)} }  \\
\label{QoS GES1}{\rm{s.t.}}\;\;&{\Upsilon _{k\overline k}^{\rm{ES}}} + {\Pi _{kk}^{\rm{ES}}} + {{\overline \gamma }_k}\sigma _k^2 \le 0,\forall k \in \left\{ {t,r} \right\},\\
\label{constraints GES1}&\eqref{amplitude GES1},\eqref{Rank W GES},\eqref{H Q W GES},\eqref{amplitude GES2},
\end{align}
\end{subequations}
\vspace{-1.2cm}

\noindent where equality constraint \eqref{DC rank} is relaxed to a penalty term added to the objective function, and $\eta  > 0$ is the penalty factor which penalizes the objective function if ${\left\{ {{{\mathbf{Q}}_k^{\rm{ES}}}} \right\}}$ is not rank-one. It can be verified that, when $\eta  \to  + \infty $, the solution ${\left\{ {{{\mathbf{Q}}_k^{\rm{ES}}}} \right\}}$ of problem \eqref{P1 GES1} always satisfies equality constraint \eqref{DC rank}, i.e., problems \eqref{P1 GES} and \eqref{P1 GES1} are equivalent~\cite{penalty2}. However, if the initial value of the penalty factor $\eta$ is chosen too large, the objective function of \eqref{P1 GES1} is dominated by the penalty term, and the impact of the desired power minimization on the solution becomes negligible. To avoid this, we first initialize $\eta$ with a small value to find a good starting point, and then, in the course of several iterations, we gradually increase $\eta$ to a sufficiently large value to eventually obtain feasible rank-one matrices. Note that, for any given penalty factor $\eta>0$, problem \eqref{P1 GES1} is still non-convex due to the non-convexity of the objective function and the non-convex constraints \eqref{QoS GES1} and \eqref{Rank W GES}. In the following, we employ SCA~\cite{SCA} to obtain a suboptimal solution of \eqref{P1 GES} in an iterative manner.\\
\indent Note that the penalty term, ${\Upsilon _{k\overline k}^{\rm{ES}}}$, and ${\Pi _{kk}^{\rm{ES}}}$ are in the form of DC functions. For a given point ${\mathbf{Q}}_k^{\left( n \right)}$ in the $n$th iteration of the SCA method, using first-order Taylor expansion, a convex upper bound for the penalty term can be obtained as follows:
\vspace{-0.4cm}
\begin{align}\label{Q uppder bound}
\begin{gathered}
  {\left\| {{{\mathbf{Q}}_k^{\rm{ES}}}} \right\|_*} - {\left\| {{{\mathbf{Q}}_k^{\rm{ES}}}} \right\|_2} \hfill \\
   \le {\left\| {{{\mathbf{Q}}_k^{\rm{ES}}}} \right\|_*} - \left\{ {{{\left\| {{\mathbf{Q}}_k^{{\rm{ES}}\left( n \right)}} \right\|}_2} + {\rm{Tr}}\left[ {\overline {\mathbf{u}} \left( {{\mathbf{Q}}_k^{{\rm{ES}}\left( n \right)}} \right){{\left( {\overline {\mathbf{u}} \left( {{\mathbf{Q}}_k^{{\rm{ES}}\left( n \right)}} \right)} \right)}^H}\left( {{{\mathbf{Q}}_k^{\rm{ES}}} - {\mathbf{Q}}_k^{{\rm{ES}}\left( n \right)}} \right)} \right]} \right\} \hfill \\
   \triangleq {\left\| {{{\mathbf{Q}}_k^{\rm{ES}}}} \right\|_*} - \overline {\mathbf{Q}} _k^{{\rm{ES}}\left( n \right)}, \hfill \\
\end{gathered}
\end{align}
\vspace{-0.6cm}

\noindent where ${\overline {\mathbf{u}} \left( {{\mathbf{Q}}_k^{{\rm{ES}}\left( n \right)}} \right)}$ denotes the eigenvector corresponding to the largest eigenvalue of ${{\mathbf{Q}}_k^{{\rm{ES}}\left( n \right)}}$, and $\overline {\mathbf{Q}} _k^{{\rm{ES}}\left( n \right)}  \triangleq {{\left\| {{\mathbf{Q}}_k^{{\rm{ES}}\left( n \right)}} \right\|}_2} + {\rm{Tr}}\left[ {\overline {\mathbf{u}} \left( {{\mathbf{Q}}_k^{{\rm{ES}}\left( n \right)}} \right){{\left( {\overline {\mathbf{u}} \left( {{\mathbf{Q}}_k^{{\rm{ES}}\left( n \right)}} \right)} \right)}^H}\left( {{{\mathbf{Q}}_k^{\rm{ES}}} - {\mathbf{Q}}_k^{{\rm{ES}}\left( n \right)}} \right)} \right]$.\\
\indent Similarly, for given points $\left\{ {{\mathbf{Q}}_k^{{\rm{ES}}\left( n \right)},{\mathbf{W}}_k^{\left( n \right)}} \right\}$ in the $n$th iteration of the SCA method, convex upper bounds of ${\Upsilon _{k\overline k}^{\rm{ES}}}$ and ${\Pi _{kk}^{\rm{ES}}}$ are respectively given by
\vspace{-0.3cm}
\begin{subequations}\label{two upper bounds}
\begin{align}\label{DC 3}
&\begin{gathered}
  {\Upsilon _{k\overline k}^{\rm{ES}}} \le \frac{{{{\overline \gamma  }_k}}}{2}\left\| {{{\mathbf{Q}}_k^{\rm{ES}}} + {{\mathbf{H}}_k}{{\mathbf{W}}_{\overline k}}{\mathbf{H}}_k^H} \right\|_F^2 + \frac{{{{\overline \gamma  }_k}}}{2}\left\| {{\mathbf{Q}}_k^{{\rm{ES}}\left( n \right)}} \right\|_F^2 - {\overline \gamma  _k}{\rm{Tr}}\left( {{{\left( {{\mathbf{Q}}_k^{{\rm{ES}}\left( n \right)}} \right)}^H}{{\mathbf{Q}}_k^{\rm{ES}}}} \right) \hfill \\
   \;\;\;\;\;\;\;\;\;\;+ \frac{{{{\overline \gamma  }_k}}}{2}\left\| {{{\mathbf{H}}_k}{\mathbf{W}}_{\overline k}^{\left( n \right)}{\mathbf{H}}_k^H} \right\|_F^2 - {\overline \gamma  _k}{\rm{Tr}}\left( {{{\left( {{\mathbf{H}}_k^H{{\mathbf{H}}_k}{\mathbf{W}}_{\overline k}^{\left( n \right)}{\mathbf{H}}_k^H{{\mathbf{H}}_k}} \right)}^H}{{\mathbf{W}}_{\overline k}}} \right) \triangleq {\left[ {{\Upsilon _{k\overline k}^{\rm{ES}}}} \right]^{ub}}, \hfill\\
\end{gathered}   \\
\label{DC 4}&\begin{gathered}
  {\Pi _{kk}^{\rm{ES}}} \le \frac{1}{2}\left\| {{{\mathbf{Q}}_k^{\rm{ES}}} - {{\mathbf{H}}_k}{{\mathbf{W}}_{k}}{\mathbf{H}}_k^H} \right\|_F^2 + \frac{1}{2}\left\| {{\mathbf{Q}}_k^{{\rm{ES}}\left( n \right)}} \right\|_F^2 - {\rm{Tr}}\left( {{{\left( {{\mathbf{Q}}_k^{{\rm{ES}}\left( n \right)}} \right)}^H}{{\mathbf{Q}}_k^{\rm{ES}}}} \right) \hfill \\
   \;\;\;\;\;\;\;\;\;\;+ \frac{1}{2}\left\| {{{\mathbf{H}}_k}{\mathbf{W}}_{k}^{\left( n \right)}{\mathbf{H}}_k^H} \right\|_F^2 - {\rm{Tr}}\left( {{{\left( {{\mathbf{H}}_k^H{{\mathbf{H}}_k}{\mathbf{W}}_{k}^{\left( n \right)}{\mathbf{H}}_k^H{{\mathbf{H}}_k}} \right)}^H}{{\mathbf{W}}_{k}}} \right) \triangleq {\left[ {{\Pi _{kk}^{\rm{ES}}}} \right]^{ub}}. \hfill \\
\end{gathered}
\end{align}
\end{subequations}
\vspace{-0.6cm}

\noindent As a result, for given points $\left\{ {{\mathbf{Q}}_k^{{\rm{ES}}\left( n \right)},{\mathbf{W}}_k^{\left( n \right)}} \right\}$, by replacing the non-convex terms with their convex upper bounds, problem \eqref{P1 GES1} is transformed into the following optimization problem:
\vspace{-0.4cm}
\begin{subequations}\label{P1 GES2}
\begin{align}
\mathop {\min }\limits_{ {{{\mathbf{W}}_k},{{\mathbf{Q}}_k^{\rm{ES}}},{{\bmm{\beta}} _k}} } &{\rm{Tr}}\left( {{{\mathbf{W}}_t}} \right) + {\rm{Tr}}\left( {{{\mathbf{W}}_r}} \right) + \eta {\sum\nolimits_{k \in \left\{ {t,r} \right\}} {\left( {{{\left\| {{{\mathbf{Q}}_k^{\rm{ES}}}} \right\|}_*} - \overline {\mathbf{Q}} _k^{{\rm{ES}}\left( n \right)}} \right)} }  \\
\label{QoS GES2}{\rm{s.t.}}\;\;&{\left[ {{\Upsilon _{k\overline k}^{\rm{ES}}}} \right]^{ub}} + {\left[ {{\Pi _{kk}^{\rm{ES}}}} \right]^{ub}} + {{\overline \gamma }_k}\sigma _k^2 \le 0,\forall k \in \left\{ {t,r} \right\},\\
\label{constraints GES2}&\eqref{amplitude GES1},\eqref{Rank W GES},\eqref{H Q W GES},\eqref{amplitude GES2}.
\end{align}
\end{subequations}
\vspace{-1.2cm}

\noindent Now, the remaining non-convexity of \eqref{P1 GES2} is the non-convex rank-one constraint \eqref{Rank W GES}. To handle this issue, we employ semidefinite relaxation (SDR) and solve the relaxed problem by ignoring \eqref{Rank W GES}. The tightness of the relaxed version of problem \eqref{P1 GES2} is shown in the following theorem.
\vspace{-0.3cm}
\begin{theorem}\label{rank one relax}
\emph{Without loss of optimality, the solutions ${\left\{ {{{\mathbf{W}}_k}} \right\}}$ obtained for the relaxed version of problem of \eqref{P1 GES2}, i.e., after dropping rank-one constraint \eqref{Rank W GES}, always satisfy ${\rm {Rank}}\left( {{{\mathbf{W}}_k}} \right) = 1,\forall k \in \left\{ {t,r} \right\}$.}
\begin{proof}
\emph{See Appendix A.}
\end{proof}
\end{theorem}
\vspace{-0.3cm}
The relaxed problem \eqref{P1 GES2}, which is a standard convex semidefinite program (SDP), can be efficiently solved via standard convex problem solvers such as CVX~\cite{cvx}. Then, we propose a penalty-based iterative algorithm for solving problem \eqref{P1 GES}, which comprises two loops. In the outer loop, the penalty factor is gradually increased from one iteration to the next as follows: $\eta  = \omega \eta $, where $\omega > 1$. The algorithm terminates when the penalty term satisfies the following criterion:
\vspace{-0.5cm}
\begin{align}\label{terminates ES}
\max \left\{ {{{\left\| {{{\mathbf{Q}}_k^{\rm{ES}}}} \right\|}_*} - {{\left\| {{{\mathbf{Q}}_k^{\rm{ES}}}} \right\|}_2},\forall k \in \left\{ {t,r} \right\}} \right\} \le {\varepsilon _1},
\end{align}
\vspace{-0.8cm}

\noindent where ${\varepsilon _1}$ denotes a predefined maximum violation of equality constraint \eqref{DC rank}. Therefore, \eqref{DC rank} will be eventually satisfied with accuracy ${\varepsilon _1}$ as $\eta$ increases. In the inner loop, $\left\{ {{\mathbf{Q}}_k^{\rm{ES}},{\mathbf{W}}_k} \right\}$ are jointly optimized by iteratively solving the relaxed version of problem \eqref{P1 GES2} for the given penalty factor. The objective function value of the relaxed version of \eqref{P1 GES2} is non-increasing in each iteration of the inner loop, and the optimal objective function value of the relaxed version of \eqref{P1 GES2} is bounded from below. Therefore, as $\eta$ approaches infinity, the developed penalty-based iterative algorithm is guaranteed to converge to a stationary point of the original problem \eqref{P1 GES}~\cite{SCA}. The details of the developed algorithm are summarized in \textbf{Algorithm 1}.\\
\begin{algorithm}[!t]\label{method1}
\caption{Proposed penalty-based iterative algorithm for solving problem \eqref{P1 GES} for ES}
\begin{algorithmic}[1]
\STATE {Initialize feasible points $\left\{ {{\mathbf{Q}}_k^{{\rm{ES}}\left( 0 \right)},{\mathbf{W}}_k^{\left( 0 \right)}} \right\}$, the penalty factor $\eta$.}
\STATE {\bf repeat: outer loop}
\STATE \quad Set iteration index $n=0$ for inner loop.
\STATE \quad {\bf repeat: inner loop}
\STATE \quad\quad For given $\left\{ {{\mathbf{Q}}_k^{{\rm{ES}}\left( n \right)},{\mathbf{W}}_k^{\left( n \right)}} \right\}$, solve the relaxed version of problem \eqref{P1 GES2}.
\STATE \quad\quad Update $\left\{ {{\mathbf{Q}}_k^{{\rm{ES}}\left( n+1 \right)},{\mathbf{W}}_k^{\left( n+1 \right)}} \right\}$ with the obtained optimal solutions, and $n=n+1$.
\STATE \quad {\bf until} the fractional decrease of the objective function value is below a predefined threshold $\epsilon_1 >0$ or the maximum number of inner iterations $n_{\max}$ is reached.
\STATE \quad Update $\left\{ {{\mathbf{Q}}_k^{{\rm{ES}}\left( 0 \right)},{\mathbf{W}}_k^{\left( 0 \right)}} \right\}$ with the current solutions $\left\{ {{\mathbf{Q}}_k^{{\rm{ES}}\left( n \right)},{\mathbf{W}}_k^{\left( n \right)}} \right\}$.
\STATE \quad Update $\eta  = \omega \eta$.
\STATE {\bf until} the constraint violation is below a predefined threshold $\varepsilon_1 >0$.
\end{algorithmic}
\end{algorithm}
\indent The computational complexity of \textbf{Algorithm 1} is analyzed as follows. The main complexity is caused by solving the relaxed version of problem \eqref{P1 GES2} in the inner loop. As the relaxed problem is a standard SDP, the computational complexity for solving this problem is ${\mathcal{O}}\left( {K{N^{3.5}} + 2{M^{3.5}}} \right)$ if the interior point method is employed~\cite{Luo}, where $K=2$ is the number of users. Then, the overall computational complexity of \textbf{Algorithm 1} is ${\mathcal{O}}\left( {{I_{{\rm{out}}}}{I_{{\rm{inn}}}}\left( {K{N^{3.5}} + 2{M^{3.5}}} \right)} \right)$, where ${I_{{\rm{inn}}}}$ and ${I_{{\rm{out}}}}$ denote the number of inner and outer iterations required for convergence, respectively. As can be observed, the computational complexity of \textbf{Algorithm 1} is polynomial in $M$, which facilitates its practical implementation even if the number of STAR-RIS elements is large.
%\subsubsection{Extension to MS} Note that for MS, it only introduces additional affine constraints \eqref{UES}, which does not affect the convexity of the resulting optimization problem. As a result, the developed penalty-based algorithm for MS is also applicable to MS by considering the linear constraint \eqref{UES}.
\subsubsection{Extended Penalty-based Algorithm for MS} Compared to ES, for MS, the formulated optimization problem in \eqref{P1 GES} involves the additional non-convex binary constraints \eqref{AC}. Therefore, we only need to focus on how to tackle this new obstacle since the other non-convex terms can be handled in a similar manner as previously described. We first transform the binary constraint \eqref{AC} equivalently into the following equality constraint:
\vspace{-0.4cm}
\begin{align}\label{integer}
{\beta _m^k} - {\left( {{\beta _m^k}} \right)^2} = 0,\forall k \in \left\{ {t,r} \right\},m \in {{\mathcal{M}}}.
\end{align}
\vspace{-1.2cm}

\noindent Recall that because of constraint \eqref{amplitude GES2} the amplitude coefficients for transmission and reflection are between 0 and 1. Thus, we always have ${\beta _m^k} - {\left( {{\beta _m^k}} \right)^2} \ge 0$, where equality holds if and only if ${\beta _m^k}$ is 0 or 1, i.e., a binary variable. As a result, the equality constraint in \eqref{integer} is satisfied only for binary variables. In the following, we extend the proposed penalty-based algorithm for ES to solve the optimization problem for MS. In particular, non-convex constraint \eqref{QoS GES}, which is highly-coupled in $\left\{ {{\mathbf{Q}}_k^{\rm{MS}},{\mathbf{W}}_k} \right\}$, and the non-convex rank-one constraint \eqref{Rank Q GES} are handled in a similar manner as previously described. Moreover, by further adding equality constraint \eqref{integer} as another penalty term into the objective function, we obtain the following optimization problem for MS:
\vspace{-0.5cm}
\begin{subequations}\label{P1 ES2}
\begin{align}
\mathop {\min }\limits_{ {{{\mathbf{W}}_k},{{\mathbf{Q}}_k^{\rm{MS}}},{{\bmm{\beta}} _k}} }& \!\!{\rm{Tr}}\!\left( {{{\mathbf{W}}_t}} \right) \!+\! {\rm{Tr}}\!\left( {{{\mathbf{W}}_r}} \right) \!+\!\eta\! \sum\nolimits_{k \in \left\{ {t,r} \right\}} \!\!{\left( {{{\left\| {{{\mathbf{Q}}_k^{\rm{MS}}}} \right\|}_*} \!\!-\! {{\left\| {{{\mathbf{Q}}_k^{\rm{MS}}}} \right\|}_2}} \right)} \!+\! \chi\! \sum\nolimits_{m = 1}^M \!{\sum\nolimits_{k \in \left\{ {t,r} \right\}} \!\!{\left( {{\beta _m^k} \!-\! {{\left( {{\beta _m^k}} \right)}\!^2}} \right)} }  \\
\label{QoS MS2}{\rm{s.t.}}\;\;&{\Upsilon _{k\overline k}^{\rm{MS}}} + {\Pi _{kk}^{\rm{MS}}} + {{\overline \gamma }_k}\sigma _k^2 \le 0,\forall k \in \left\{ {t,r} \right\},\\
\label{constraints MS2}{\rm{s.t.}}\;\;&\eqref{amplitude GES1},\eqref{Rank W GES},\eqref{H Q W GES},\eqref{amplitude GES2},
\end{align}
\end{subequations}
\vspace{-1.2cm}

\noindent where $\left\{ {\Upsilon _{k\overline k}^{{\rm{MS}}},\Pi _{kk}^{{\rm{MS}}},\forall k \in \left\{ {t,r} \right\}} \right\}$ are obtained by replacing ${{\mathbf{Q}}_k^{\rm{ES}}}$ in \eqref{DC 1} and \eqref{DC 2} by ${{\mathbf{Q}}_k^{\rm{MS}}}$, and $\chi>0$ is a new penalty factor which penalizes the objective function if $\left\{ {{\beta _m^k}} \right\}$ belongs to $\left( {0,1} \right)$. Similarly, it can be verified that, when $\eta, \chi  \to  + \infty $, the solution obtained from problem \eqref{P1 ES2} satisfies equality constraints \eqref{DC rank} and \eqref{integer}. As problem \eqref{P1 ES2} has a similar structure as problem \eqref{P1 GES1}, we can still employ SCA to solve this new non-convex optimization problem. For given points $\left\{{{\bm{\beta}} _k^{\left( n \right)}} \right\}$ in the $n$th iteration of the SCA method, by using first-order Taylor expansion, an upper bound for the new penalty term can be obtained as follows:
\vspace{-0.3cm}
\begin{align}\label{ub binary}
\begin{gathered}
  \beta _m^k - {\left( {\beta _m^k} \right)^2} \le \beta _m^k - {\left( {\beta _m^{k\left( n \right)}} \right)^2} - 2\beta _m^{k\left( n \right)}\left( {\beta _m^k - \beta _m^{k\left( n \right)}} \right) \hfill \\
  \;\;\;\;\;\;\;\;\;\;\;\;\;\;\;\;\;\;\; = \left( {1 - 2\beta _m^{k\left( n \right)}} \right)\beta _m^k + {\left( {\beta _m^{k\left( n \right)}} \right)^2} \triangleq \Omega \left( {\beta _m^k,\beta _m^{k\left( n \right)}} \right),\forall k \in \left\{ {t,r} \right\},m \in {{\mathcal{M}}}. \hfill \\
\end{gathered}
\end{align}
\vspace{-0.8cm}

\noindent By replacing the non-convex terms with their upper bounds for given points $\left\{ {{\mathbf{Q}}_k^{\left( n \right)},{\mathbf{W}}_k^{\left( n \right)}}, {{\bm{\beta}} _k^{\left( n \right)}} \right\}$, problem \eqref{P1 ES2} can be reformulated as follows:
\vspace{-0.5cm}
\begin{subequations}\label{ES2 ap}
\begin{align}
\mathop {\min }\limits_{ {{{\mathbf{W}}_k},{{\mathbf{Q}}_k^{\rm{MS}}},{\bmm{\beta} _k}} } &\!\!{\rm{Tr}}\!\left( {{{\mathbf{W}}_t}} \right) \!+\! {\rm{Tr}}\!\left( {{{\mathbf{W}}_r}} \right) \!+ \eta\! \sum\nolimits_{k \in \left\{ {t,r} \right\}}\!\! {\left(\! {{{\left\| {{{\mathbf{Q}}_k^{\rm{MS}}}} \right\|}_*} \!-\! \overline {\mathbf{Q}} _k^{{\rm{MS}}\left( n \right)}} \!\right)}  \!  + \!\chi\! \sum\nolimits_{m = 1}^M\! {\sum\nolimits_{k \in \left\{ {t,r} \right\}}\!\Omega \left( {\beta _m^k,\beta _m^{k\left( n \right)}} \right) }   \\
\label{QoS MS}{\rm{s.t.}}\;\;&{\left[ {{\Upsilon _{k\overline k}^{\rm{MS}}}} \right]^{ub}} + {\left[ {{\Pi _{kk}^{\rm{MS}}}} \right]^{ub}} + {{\overline \gamma }_k}\sigma _k^2 \le 0,\forall k \in \left\{ {t,r} \right\},\\
\label{constraints ES2 ap}&\eqref{amplitude GES1},\eqref{Rank W GES},\eqref{H Q W GES},\eqref{amplitude GES2},
\end{align}
\end{subequations}
\vspace{-0.8cm}

\noindent where the remaining non-convexity is caused by rank-one constraint \eqref{Rank W GES}. We can employ again SDR and remove the non-convex rank-one constraint. The tightness of the relaxation can be proved similar to \textbf{Theorem 1}. As a result, the relaxed version of problem \eqref{ES2 ap} is convex and can be efficiently solved via standard convex problem solvers such as CVX~\cite{cvx}. Similar to \textbf{Algorithm 1} for ES, for MS, a two-loop penalty-based iterative algorithm for solving problem \eqref{P1 GES} can be developed. This algorithm is summarized in \textbf{Algorithm 2}. Since for MS there are two penalty terms, the termination criterion in the outer loop is given by
\vspace{-0.3cm}
\begin{align}\label{terminal MS}
\max \left\{ {{{\left\| {{{\mathbf{Q}}_k^{\rm{MS}}}} \right\|}_*} - {{\left\| {{{\mathbf{Q}}_k^{\rm{MS}}}} \right\|}_2},{\beta _m^k} - {{\left( {{\beta _m^k}} \right)}^2},\forall k \in \left\{ {t,r} \right\},m \in {{\mathcal{M}}}} \right\} \le {\varepsilon _2},
\end{align}
\vspace{-1.2cm}

\noindent where ${\varepsilon _2} > 0$ is the predefined accuracy with which equality constraints \eqref{DC rank} and \eqref{integer} are met.
\begin{algorithm}[!t]\label{method2}
\caption{Proposed penalty-based iterative algorithm for solving problem \eqref{P1 GES} for MS}
\begin{algorithmic}[1]
\STATE {Initialize feasible points $\left\{ {{\mathbf{Q}}_k^{{\rm{MS}}\left( 0 \right)},{\mathbf{W}}_k^{\left( 0 \right)}}, {{\bm{\beta}} _k^{\left( 0 \right)}} \right\}$, the penalty factors $\eta$ and $\chi$.}
\STATE {\bf repeat: outer loop}
\STATE \quad Set iteration index $n=0$ for inner loop.
\STATE \quad {\bf repeat: inner loop}
\STATE \quad\quad For given $\left\{ {{\mathbf{Q}}_k^{{\rm{MS}}\left( n \right)},{\mathbf{W}}_k^{\left( n \right)}}, {{\bm{\beta}} _k^{\left( n \right)}} \right\}$, solve the relaxed version of problem \eqref{ES2 ap}.
\STATE \quad\quad Update $\left\{ {{\mathbf{Q}}_k^{{\rm{MS}}\left( n+1 \right)},{\mathbf{W}}_k^{\left( n+1 \right)}}, {{\bm{\beta}} _k^{\left( n+1 \right)}} \right\}$ with the obtained optimal solutions, and $n=n+1$.
\STATE \quad {\bf until} the fractional decrease of the objective function value is below a predefined threshold $\epsilon_2 >0$ or the maximum number of inner iterations $n_{\max}$ is reached.
\STATE \quad Update $\left\{ {{\mathbf{Q}}_k^{{\rm{MS}\left( 0 \right)}},{\mathbf{W}}_k^{\left( 0 \right)}}, {{\bm{\beta}} _k^{\left( 0 \right)}} \right\}$ with the current solutions $\left\{ {{\mathbf{Q}}_k^{{\rm{MS}}\left( n \right)},{\mathbf{W}}_k^{\left( n \right)}}, {{\bm{\beta}} _k^{\left( n \right)}} \right\}$.
\STATE \quad Update $\eta  = \omega \eta $, $\chi  = \varpi  \chi $.
\STATE {\bf until} the constraint violation is below a predefined threshold $\varepsilon_2 >0$.
\end{algorithmic}
\end{algorithm}
\vspace{-0.2cm}
\begin{remark}\label{multiuser ES MS}
\emph{Although, in this paper, we focus on systems with one T user and one R user, the proposed \textbf{Algorithms 1} and \textbf{2} can be extended to ES/MS STAR-RIS aided communication systems with multiple T and R users. Suppose that there are $J>2$ users.  The corresponding optimization problem can be obtained from optimization problem \eqref{P1 GES} by replacing the inter-user interference in the first term of \eqref{QoS GES} (i.e., ${\overline \gamma  _k}{\rm{Tr}}\left( {{{\mathbf{Q}}_k^{\rm{Y}}}{{\mathbf{H}}_k}{{\mathbf{W}}_{\overline k }}{\mathbf{H}}_k^H} \right)$) with the sum of the inter-user interference caused by all $J-1$ interfering users and minimizing the power consumption caused by the $J$ active beamforming vectors. In particular, the $J-1$ new non-convex inter-user interference terms can be tackled in a similar manner as previously discussed (i.e., \eqref{DC 1} and \eqref{DC 3}). Therefore, the optimization problem for ES and MS for multiple T and R users can still be solved with the proposed \textbf{Algorithms 1} and \textbf{2}.}
\end{remark}
\vspace{-0.8cm}
\subsection{Proposed Solution for TS}
\vspace{-0.2cm}
To solve the optimization problem for TS, let ${\mathbf{q}}_k^{\rm{TS}} = {\left[ {{e^{j\theta _1^k}},{e^{j\theta _2^k}}, \ldots ,{e^{j\theta _M^k}}} \right]^H},\forall k \in \left\{ {t,r} \right\}$, denote the STAR-RIS transmission/reflection-coefficient vector for $0 \le{\lambda ^k}\le 1$ fraction of the total communication time. The effective channel power gain of user $k$ is given by ${\left| {{\mathbf{v}}_k^H{\mathbf{\Theta }}_k^{\rm{TS}}{\mathbf{G}}{{\mathbf{w}}_k}} \right|^2} = {\left| {{{\left( {{\mathbf{q}}_k^{\rm{TS}}} \right)}^H}{{\mathbf{H}}_k}{{\mathbf{w}}_k}} \right|^2}$, where ${{\mathbf{H}}_k} = {\rm{diag}}\left( {{\mathbf{v}}_k^H} \right){\mathbf{G}},\forall k \in \left\{ {t,r} \right\}$. Then, for TS, optimization problem \eqref{P1 X} can be rewritten as follows:
\vspace{-0.4cm}
\begin{subequations}\label{P1 TS}
\begin{align}
&\mathop {\min }\limits_{ {{{\mathbf{w}}_k},{{\mathbf{q}}_k^{\rm{TS}}},{\lambda ^k}} } \;\;{\left\| {{{\mathbf{w}}_t}} \right\|^2} + {\left\| {{{\mathbf{w}}_r}} \right\|^2} \\
\label{QoS TS}{\rm{s.t.}}\;\;& {\lambda ^k}{\log _2}\left( {1 + \frac{{{{\left| {{{\left( {{\mathbf{q}}_k^{\rm{TS}}} \right)}^H}{{\mathbf{H}}_k}{{\mathbf{w}}_k}} \right|}^2}}}{{{\lambda ^k}\sigma _k^2}}} \right) \ge {{\overline R }_k},\forall k \in \left\{ {t,r} \right\},\\
\label{Theta TS}&{\left| {{{\left[ {{{\mathbf{q}}_k^{\rm{TS}}}} \right]}_m}} \right|^2} = 1,,\forall k \in \left\{ {t,r} \right\}, m \in {{\mathcal{M}}},\\
\label{lambda TS}&0 \le {\lambda ^t} \le 1,0 \le {\lambda ^r} \le 1,{\lambda ^t} + {\lambda ^r} = 1.
\end{align}
\end{subequations}
\vspace{-1.3cm}

\noindent As, in each time instant, only one user receives data from the BS via the STAR-RIS, for any given transmission/reflection-coefficient vector, the optimal active beamforming vector at the BS is the maximum-ratio transmission (MRT) beamformer~\cite{Wu2019IRS}, i.e., ${\mathbf{w}}_k^* = \sqrt {{p_k}} \frac{{{\mathbf{H}}_k^H{\mathbf{q}}_k^{\rm{TS}}}}{{{{\left\| {{{\left( {{\mathbf{q}}_k^{\rm{TS}}} \right)}^H}{{\mathbf{H}}_k}} \right\|}_2}}}$, where ${{p_k}}$ denotes the allocated transmit power for user $k$. By substituting $\left\{ {{\mathbf{w}}_k^*} \right\}$ into \eqref{P1 TS}, we obtain the following problem:
\vspace{-0.8cm}
\begin{subequations}\label{P1 TS1}
\begin{align}
&\mathop {\min }\limits_{ {{p_k},{{\mathbf{q}}_k^{\rm{TS}}},{\lambda ^k}} } \;\;{p_t} + {p_r} \\
\label{QoS TS1}{\rm{s.t.}}\;\;& {\lambda ^k}{\log _2}\left( {1 + \frac{{{p_k}}{\left\| {{{\left( {{\mathbf{q}}_k^{\rm{TS}}} \right)}^H}{{\mathbf{H}}_k}} \right\|^2}}{{{\lambda ^k}\sigma _k^2}}} \right) \ge {{\overline R }_k},\forall k \in \left\{ {t,r} \right\}\\
\label{Theta TS1}&\eqref{Theta TS},\eqref{lambda TS}.
\end{align}
\end{subequations}
\vspace{-1.3cm}

\noindent Since the two users are alternatingly served, the optimal transmission/reflection-coefficient vectors for problem \eqref{P1 TS1} maximize the effective channel gain of each user, which yields the following subproblem:
\vspace{-0.6cm}
\begin{subequations}\label{P1 TS1 sub}
\begin{align}
{\mathbf{q}}_k^{\rm{TS}*} =& \arg \max \;\;{{\left( {{\mathbf{q}}_k^{\rm{TS}}} \right)}^H}{{\mathbf{H}}_k}{\mathbf{H}}_k^H{{\mathbf{q}}_k^{\rm{TS}}},\forall k \in \left\{ {t,r} \right\}\\
\label{um}{\rm{s.t.}}\;\;&{\left| {{{\left[ {{{\mathbf{q}}_k^{\rm{TS}}}} \right]}_m}} \right|^2} = 1,\forall m \in {{\mathcal{M}}}.
\end{align}
\end{subequations}
\vspace{-1.2cm}

\noindent The above transmission/reflection-coefficient design problem can be efficiently solved using either the suboptimal low-complexity SDR based algorithm proposed in \cite{Wu2019IRS} or the globally optimal branch-and-bound based algorithm proposed in \cite{Yu_optimal}. The details are omit here for brevity.\\
\indent With the obtained desired solutions for the transmission/reflection-coefficient vectors, $\left\{ {{\mathbf{q}}_k^{\rm{TS}*}} \right\}$, problem \eqref{P1 TS1} is reduced to the following subproblem:
\vspace{-0.6cm}
\begin{subequations}\label{P1 TS1 sub2}
\begin{align}
&\mathop {\min }\limits_{ {{p_k},{\lambda ^k}} } \;\;{p_t} + {p_r} \\
\label{QoS TS12}{\rm{s.t.}}\;\;& {\lambda ^k}{\log _2}\left( {1 + \frac{{{p_k}}{\left\| {{{\left( {{\mathbf{q}}_k^{\rm{TS}*}} \right)}^H}{{\mathbf{H}}_k}} \right\|^2}}{{{\lambda ^k}\sigma _k^2}}} \right) \ge {{\overline R }_k},\forall k \in \left\{ {t,r} \right\}\\
\label{Theta TS12}&\eqref{lambda TS}.
\end{align}
\end{subequations}
\vspace{-0.8cm}

\noindent The above resource allocation problem is a convex optimization problem, since the LHS of \eqref{QoS TS12} is a jointly concave function with respect to ${{p_k}}$ and ${\lambda ^k}$, and \eqref{lambda TS} is an affine constraint. Therefore, problem \eqref{P1 TS1 sub2} can be efficiently solved using standard convex problem solvers such as CVX~\cite{cvx}. Based on the above two subproblems, the procedure for solving problem \eqref{P1 TS} for TS is summarized in \textbf{Algorithm 3}. The complexity of \textbf{Algorithm 3} is analyzed as follows. Suppose that the suboptimal SDR based algorithm \cite{Wu2019IRS} is employed for solving problem \eqref{P1 TS1 sub} for each user. Then, the computational complexity is ${\mathcal{O}}\left( {\left( {K{M^{3.5}}} \right)} \right)$~\cite{Wu2019IRS,Luo}, where $K=2$ is the number of users. For the convex resource allocation problem \eqref{P1 TS1 sub2}, the computational complexity is ${{\mathcal{O}}}\left( {{{\left( {2K} \right)}^{3.5}}} \right)$~\cite{convex}. Therefore, the overall computational complexity of \textbf{Algorithm 3} is ${{\mathcal{O}}}\left( {K{M^{3.5}} + {{\left( {2K} \right)}^{3.5}}} \right)$. \textbf{Algorithm 3} for TS requires a lower computational complexity than \textbf{Algorithms 1} and \textbf{2} for ES and MS, since transmission and reflection are decoupled for TS.
\begin{algorithm}[!t]\label{method3}
\caption{Procedure for solving problem \eqref{P1 TS} for TS}
\begin{algorithmic}[1]
\STATE {Obtain the transmission/reflection-coefficient vectors, denoted by $\left\{ {{\mathbf{q}}_k^{\rm{TS}*}}, \forall k \in \left\{ {t,r} \right\}\right\}$, by solving the corresponding effective channel gain maximization problem \eqref{P1 TS1 sub}.}
\STATE {For given $\left\{ {{\mathbf{q}}_k^{\rm{TS}*}}, \forall k \in \left\{ {t,r} \right\}\right\}$, solve the convex resource allocation problem \eqref{P1 TS1 sub2} to obtain the optimal solutions $\left\{ {p_k^*,{\lambda ^{k*}}, \forall k \in \left\{ {t,r} \right\}} \right\}$.}
\end{algorithmic}
\end{algorithm}
\vspace{-0.3cm}
\begin{remark}\label{multiuser TS}
\emph{Since the proposed TS protocol severs only one user in each time instant, for systems with $J>2$ users, the proposed \textbf{Algorithm 3} can be directly applied as follows. First, solve each effective channel gain maximization problem \eqref{P1 TS1 sub} to obtain $J$ transmission/reflection-coefficient vectors. Then, solve the resulting convex resource allocation problem \eqref{P1 TS1 sub2} with $J$ power allocation and time allocation variables, respectively.}
\end{remark}
\vspace{-0.8cm}
\section{Numerical Results}
\begin{figure}[t!]
  \centering
  \includegraphics[width=3in]{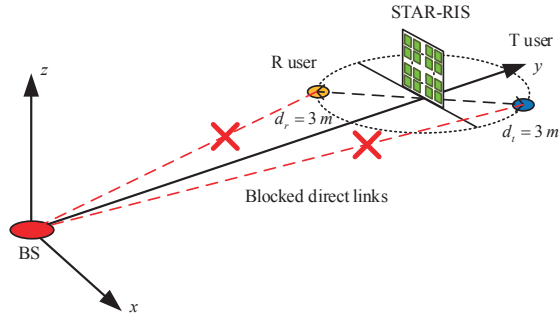}\\
  \caption{The simulated setup.}\label{setup}
\end{figure}
In this section, numerical results are provided to validate the effectiveness of the proposed STAR-RIS aided communication system.
\vspace{-0.6cm}
\subsection{Simulation Setup}
\vspace{-0.2cm}
Fig. \ref{setup} illustrates the considered three-dimensional (3D) simulation setup, where the BS and the STAR-RIS\footnote{In practice, both the horizontal location and the height of the STAR-RIS can be either optimized or selected based on the environmental conditions.} are located at $\left( {0,0,0} \right)$ meters and $\left( {0,50,0} \right)$ meters, respectively. The T and R users are randomly located on half-circles centered at the STAR-RIS with a radius of ${d_t} = {d_r} = 3$ m. The STAR-IRS is assumed to be equipped with a uniform planar array (UPA) composed of $M = {M_h}{M_v}$ elements, where ${M_h}=5$ denotes the number of elements along the horizontal plane and we increase ${M_v}$ linearly with $M$. For our simulations, the narrow-band quasi-static fading channels from the BS to the STAR-RIS and from the STAR-RIS to the two users are modeled as Rician fading channels as follows:
\vspace{-0.3cm}
\begin{subequations}
\begin{align}\label{AP RIS channel}
{\mathbf{G}}& = \sqrt {\frac{{{\rho _0}}}{{{{d_{BR}^{{\alpha _{BR}}}}}}}} \left( {\sqrt {\frac{{{K_{BR}}}}{{{K_{BR}} + 1}}} {\mathbf{G}}^{{\rm{LoS}}} + \sqrt {\frac{1}{{{K_{BR}} + 1}}} {\mathbf{G}}^{{\rm{NLoS}}}} \right),\\
\label{RIS-USER channel}
{{\mathbf{v}}_k} &= \sqrt {\frac{{{\rho _0}}}{{d_{RU,k}^{{\alpha _{RU}}}}}} \left( {\sqrt {\frac{{{K_{RU}}}}{{{K_{RU}} + 1}}} {\mathbf{v}}_k^{{\rm{LoS}}} + \sqrt {\frac{1}{{{K_{RU}} + 1}}} {\mathbf{v}}_k^{{\rm{NLoS}}}} \right),\forall k \in \left\{ {t,r} \right\},
\end{align}
\end{subequations}
\vspace{-1.0cm}

\noindent where ${d_{BR}}$ and ${d_{RU,k}}$ denote the distances between the BS and the STAR-RIS and between the STAR-RIS and user $k$, ${{\alpha _{BR}}}$ and ${{\alpha _{RU}}}$ denote the corresponding path loss exponents, $K_{BR}$ and $K_{RU}$ denote the Rician factors, ${\rho _0}$ represents the path loss at a reference distance of 1 meter, ${{\mathbf{G}}^{{\rm{LoS}}}}$ and ${{\mathbf{v}}_k^{{\rm{LoS}}}}$ are the deterministic line-of-sight (LoS) components, and ${{\mathbf{G}}^{{\rm{NLoS}}}}$ and ${{\mathbf{v}}_k^{{\rm{NLoS}}}}$ are the random non-line-of-sight (NLoS) components modeled as Rayleigh fading. The adopted system parameters are presented in Table \ref{Parameters}. Without loss of generality, we assume that the two users for unicast transmission have the same QoS requirements, i.e., ${\overline R _t} = {\overline R _r} = {\overline R _0} \triangleq {\log _2}\left( {1 + {{\overline \gamma }_0}} \right)$ and ${{\overline \gamma }_0}$ is the minimum required signal-to-interference-plus-noise ratio (SINR). The communication requirement for multicast transmission is set to ${\overline R _c} \triangleq {\log _2}\left( {1 + {{\overline \gamma }_c}} \right)$, where ${{\overline \gamma }_c}$ is the minimum required SINR. The power consumption results shown (i.e., Figs. \ref{CvM}-\ref{CvR}) were obtained by averaging over 100 channel realizations. Specifically, we first randomly generated 100 user distributions, and then obtained the corresponding channel realizations.
\vspace{-0.4cm}
\begin{table*}[h!]
\caption{System Parameters}
\vspace{-0.6cm}
\begin{center}
\centering
\resizebox{\textwidth}{!}{
\begin{tabular}{|l|l|l||l|l|l|}
\hline
\centering
${{\alpha _{BR}}}$  & Path loss exponent for BS-STAR-RIS channels &$2.2$ & ${{\alpha _{RU}}}$ &Path loss exponent for STAR-RIS-user channels&$2.2$\\
\hline
\centering
$K_{BR}$ &  Rician factor for BS-STAR-RIS channels  & $3$ dB & $K_{BU}$ &Rician factor for STAR-RIS-user channels&$3$ dB\\
\hline
\centering
${\rho _0}$&  Path loss at 1 meter  & $-30$ dB &$\sigma _k^2$  &User noise power&$-90$ dBm \\
\hline
\centering
$\eta$, $\chi$& Initialized penalty factors for Algorithms 1 and 2  & ${10^{ - 4}}$   & $\omega$, $\varpi$ &Scaling factors for Algorithms 1 and 2 &$10$\\
\hline
\centering
${{\epsilon}_1}$, ${{\epsilon}_2}$& Convergence tolerance for SCA  & ${10^{ - 2}}$   & ${\varepsilon _1}$, ${\varepsilon _2}$ &Accuracy for equality constraints &${10^{ - 7}}$\\
\hline
\centering
$n_{\max}$ & Maximum number of inner iterations of Algorithms 1 and 2  & $30$   &  & &\\
\hline
\end{tabular}
}
\end{center}
\label{Parameters}
\end{table*}%105
%set as follows:  ${{\alpha _{BR}}}={{\alpha _{RU}}}=2.2$, $K_{BR}=K_{RU}=3$ dB, ${\rho _0}=-30$ dB, and $\sigma _k^2 =  - 90$ dBm. For the proposed penalty-based \textbf{Algorithms 1} and \textbf{2}, the parameters are set as follows: $\eta  = \chi  = {10^{ - 4}}$, $\omega  = \varpi  = 10$, ${{\epsilon}_1} = {{\epsilon}_2} = {10^{ - 2}}$, ${\varepsilon _1} = {\varepsilon _2} = {10^{ - 7}}$, and $n_{\max}=30$.
\vspace{-1.6cm}
\subsection{Baseline Schemes}
\vspace{-0.2cm}
To verify the effectiveness of the proposed STAR-RIS concept and the corresponding operating protocols, we compare with the following two baseline schemes.
\begin{itemize}
  \item \textbf{Baseline scheme 1 (also referred to as conventional RISs)}: In this case, the full-space coverage facilitated by the STAR-RIS in Fig. \ref{setup} is achieved by employing one conventional reflecting-only RIS and one transmitting-only RIS. The two conventional RISs are deployed adjacent to each other at the same location as the STAR-RIS. For a fair comparison, each conventional reflecting/transmitting RIS is assumed to have ${M \mathord{\left/
 {\vphantom {M 2}} \right.
 \kern-\nulldelimiterspace} 2}$ elements, where $M$ is assumed to be an even number for simplicity. This baseline scheme can be regarded as a special case of an MS STAR-RIS, where ${M \mathord{\left/
 {\vphantom {M 2}} \right.
 \kern-\nulldelimiterspace} 2}$ elements operate in the T mode and ${M \mathord{\left/
 {\vphantom {M 2}} \right.
 \kern-\nulldelimiterspace} 2}$ elements operate in the R mode. Therefore, the resulting optimization problem can be solved by applying \textbf{Algorithm 1} with ${\bm{\beta} ^t} = [{{\mathbf{1}}_{1 \times {M \mathord{\left/
 {\vphantom {M 2}} \right.
 \kern-\nulldelimiterspace} 2}}}\;{{\mathbf{0}}_{{{1 \times M} \mathord{\left/
 {\vphantom {{1 \times M} 2}} \right.
 \kern-\nulldelimiterspace} 2}}}]$ and ${\bm{\beta} ^r}  = [{{\mathbf{0}}_{{{1 \times M} \mathord{\left/
 {\vphantom {{1 \times M} 2}} \right.
 \kern-\nulldelimiterspace} 2}}}\;{{\mathbf{1}}_{1 \times {M \mathord{\left/
 {\vphantom {M 2}} \right.
 \kern-\nulldelimiterspace} 2}}}]$.
  \item \textbf{Baseline scheme 2 (also referred to as uniform energy splitting (UES))}: In this case, we assume that all elements of the ES STAR-RIS employ the same amplitude coefficients for transmission and reflection, respectively, i.e., $\beta _m^k = {\overline \beta  ^k},\beta _m^r = {\overline \beta  ^r}\forall m \in {\mathcal{M}}$, where $0 \le {\overline \beta  ^k},{\overline \beta  ^r} \le 1$ and ${\overline \beta  ^t} + {\overline \beta  ^r} = 1$. UES can be regarded as a special case of the ES STAR-RIS employing a group/surface-wise amplitude design. The resulting optimization problem can be solved by applying \textbf{Algorithm 1} with the above linear equality constraints.
\end{itemize}
\vspace{-0.6cm}
\subsection{Convergence of Algorithms 1 and 2}
\begin{figure}[t!]
\centering
\subfigure[Power consumption.]{\label{Power_UMC}
\includegraphics[width= 2.0in]{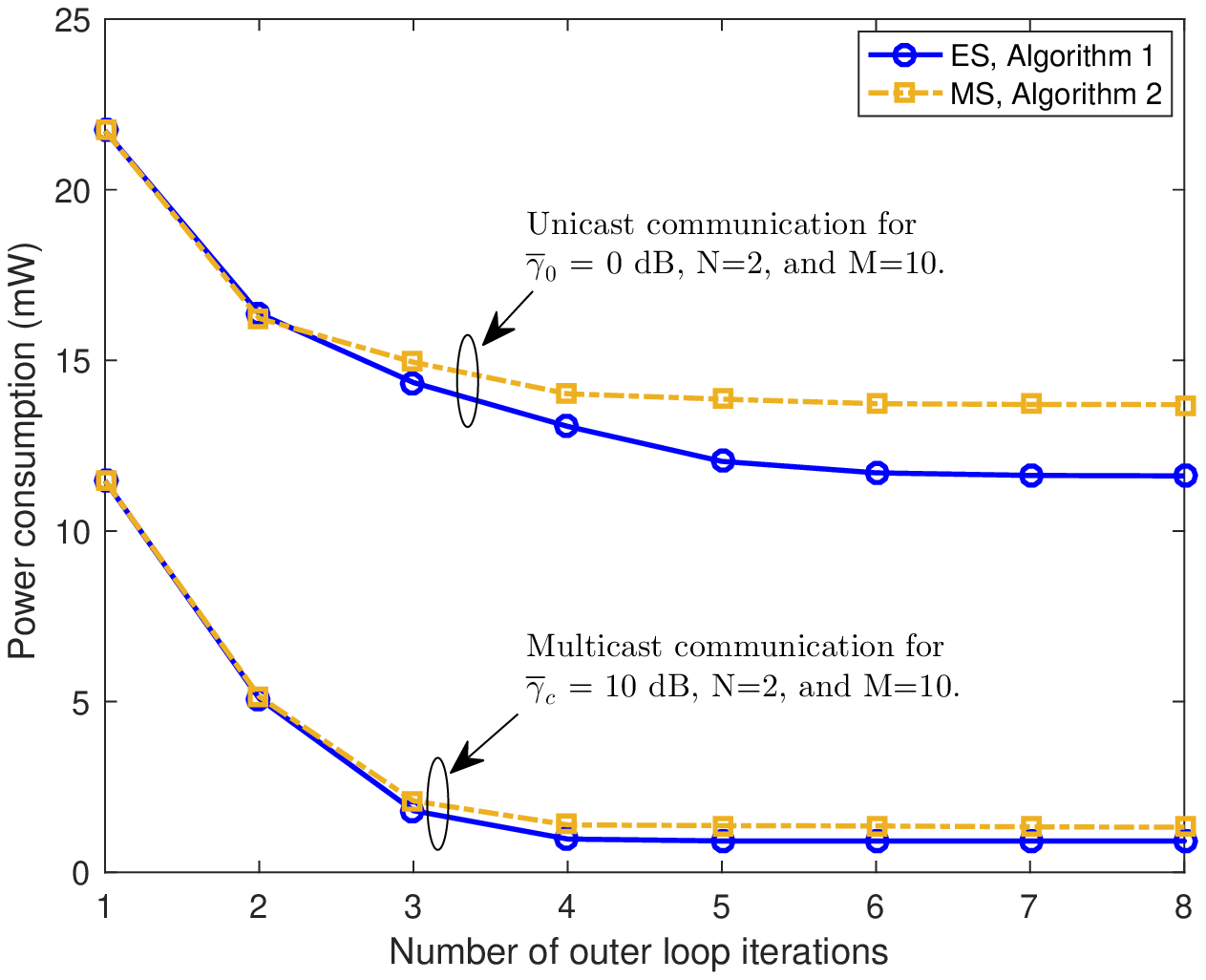}}
\subfigure[Constraint violation for unicast.]{\label{Constraint_UC}
\includegraphics[width= 2.0in]{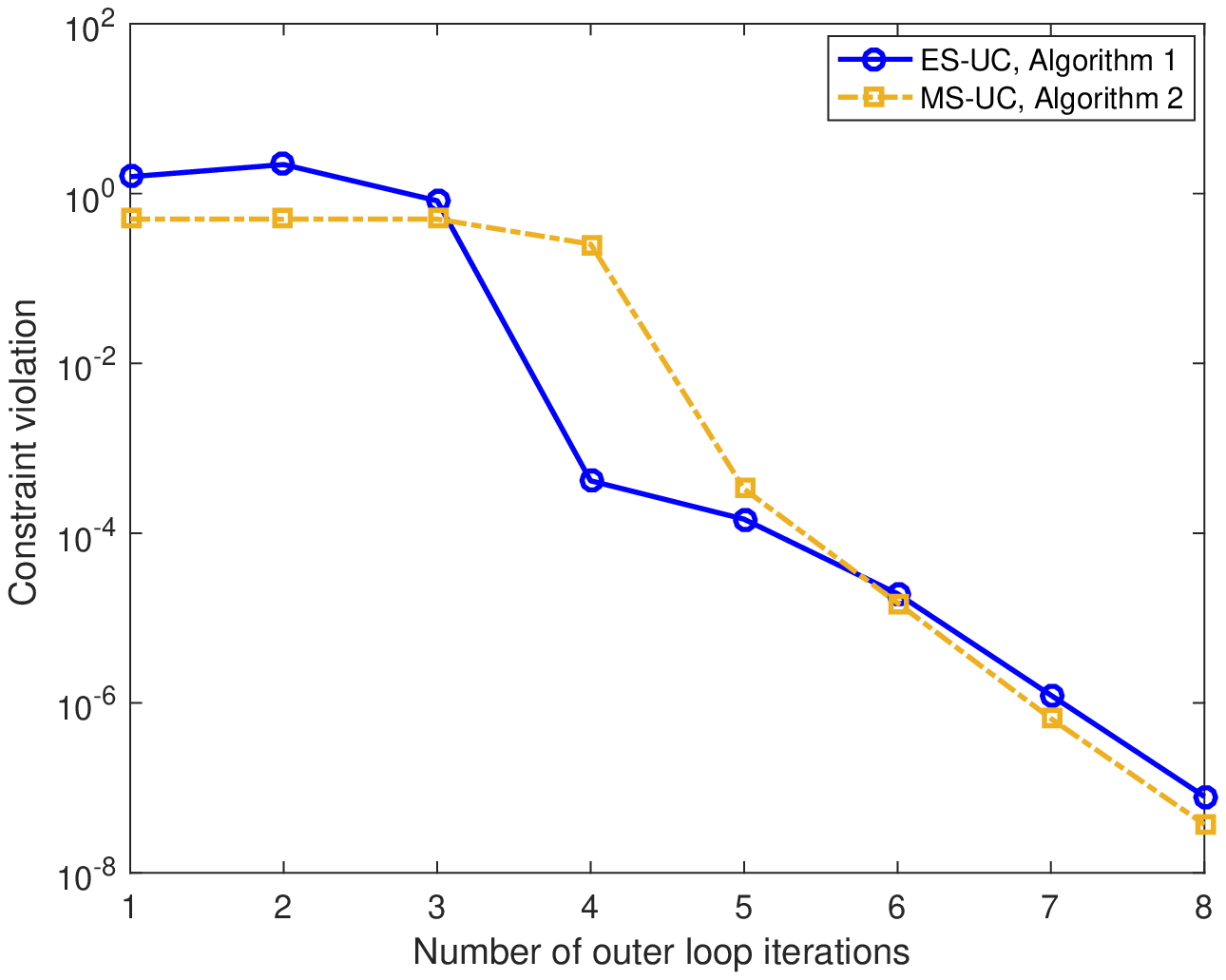}}
\subfigure[Constraint violation for multicast.]{\label{Constraint_MC}
\includegraphics[width= 2.0in]{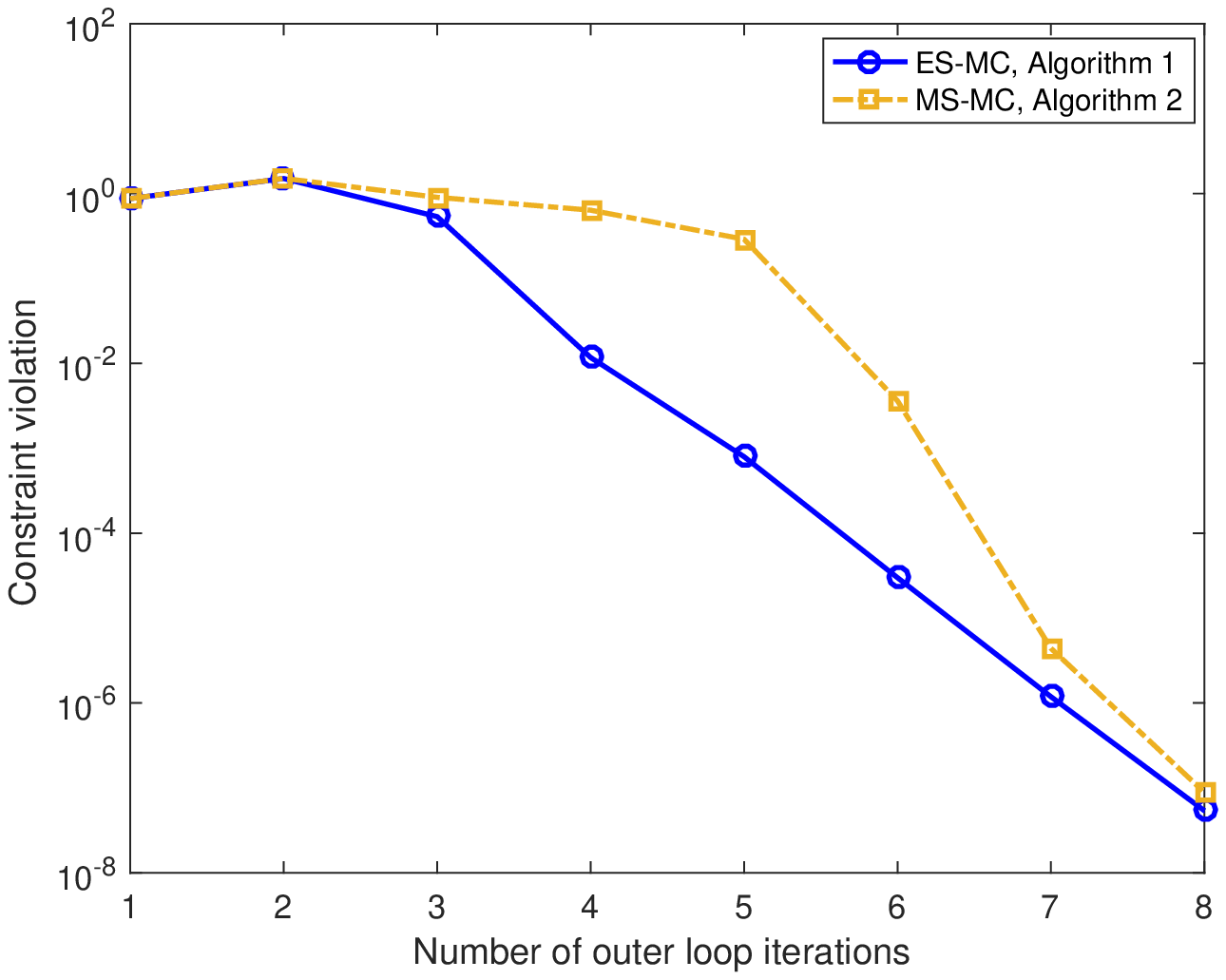}}
\caption{Convergence behaviour of proposed penalty-based iterative algorithms, where ``UC'' and ``MC'' represent unicast and multicast, respectively.}\label{Convergence}
\end{figure}

\vspace{-0.2cm}
In Fig. \ref{Convergence}, we investigate the convergence behavior and the violation of the equality constraints, i.e., \eqref{terminates ES} and \eqref{terminal MS} of the proposed \textbf{Algorithm 1} for ES and \textbf{Algorithm 2} for MS. Both unicast and multicast transmission are considered. We set $N=2$, $M=10$, and $\overline \gamma_0 = 0$ dB for unicast communication, and $\overline \gamma_c = 10$ dB for multicast communication. The presented results were obtained for one random channel realization. As shown in Fig. \ref{Power_UMC}, the required power consumption obtained for \textbf{Algorithms 1} and \textbf{2} decreases quickly as the number of outer loop iterations increases. Specifically, the proposed penalty-based algorithms converge in 6 iterations and 4 iterations for unicast and multicast communication, respectively. This is because unicast communication involves more optimization variables (i.e., two different active beamforming vectors at the BS) and more complex constraints (i.e., inter-user interference terms) than multicast communication, and thus more iterations are needed for convergence. Fig. \ref{Constraint_UC} and Fig. \ref{Constraint_MC} illustrate the constraint violation for \textbf{Algorithms 1} and \textbf{2} versus the number of outer loop iterations for unicast and multicast communication, respectively. As can be observed, the constraint violation in all setups decreases quickly as the number of outer loop iterations increases, and ultimately reaches the predefined accuracy (i.e., ${\varepsilon _1} = {\varepsilon _2} = {10^{ - 7}}$) after 8 iterations. This means that feasible rank-one transmission/reflection-coefficient matrices, $\left\{ {{\mathbf{Q}}_k^{\rm{ES/MS}}} \right\}$, and binary transmission/reflection amplitude coefficients, $\left\{ {\beta _m^k} \right\}$, are obtained with \textbf{Algorithms 1} and \textbf{2}.
\vspace{-0.6cm}
\subsection{Required Power Consumption Versus Number of RIS Elements}
\begin{figure}[t!]
\centering
\subfigure[Unicast communication, $\overline \gamma _0$ = 0 dB.]{\label{UCvM}
\includegraphics[width= 3in]{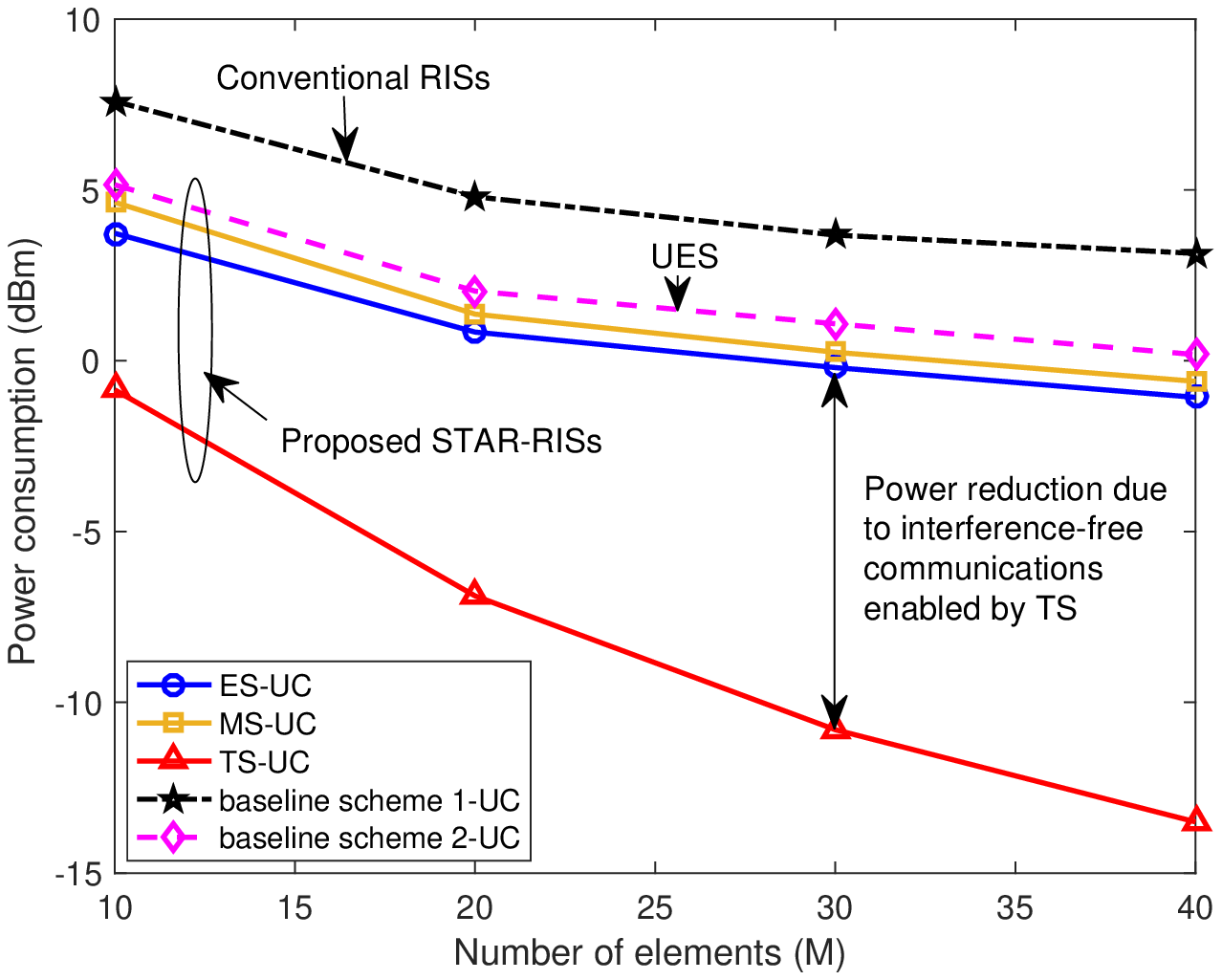}}
\subfigure[Multicast communication, $\overline \gamma _c$ = 10 dB.]{\label{MCvM}
\includegraphics[width= 3in]{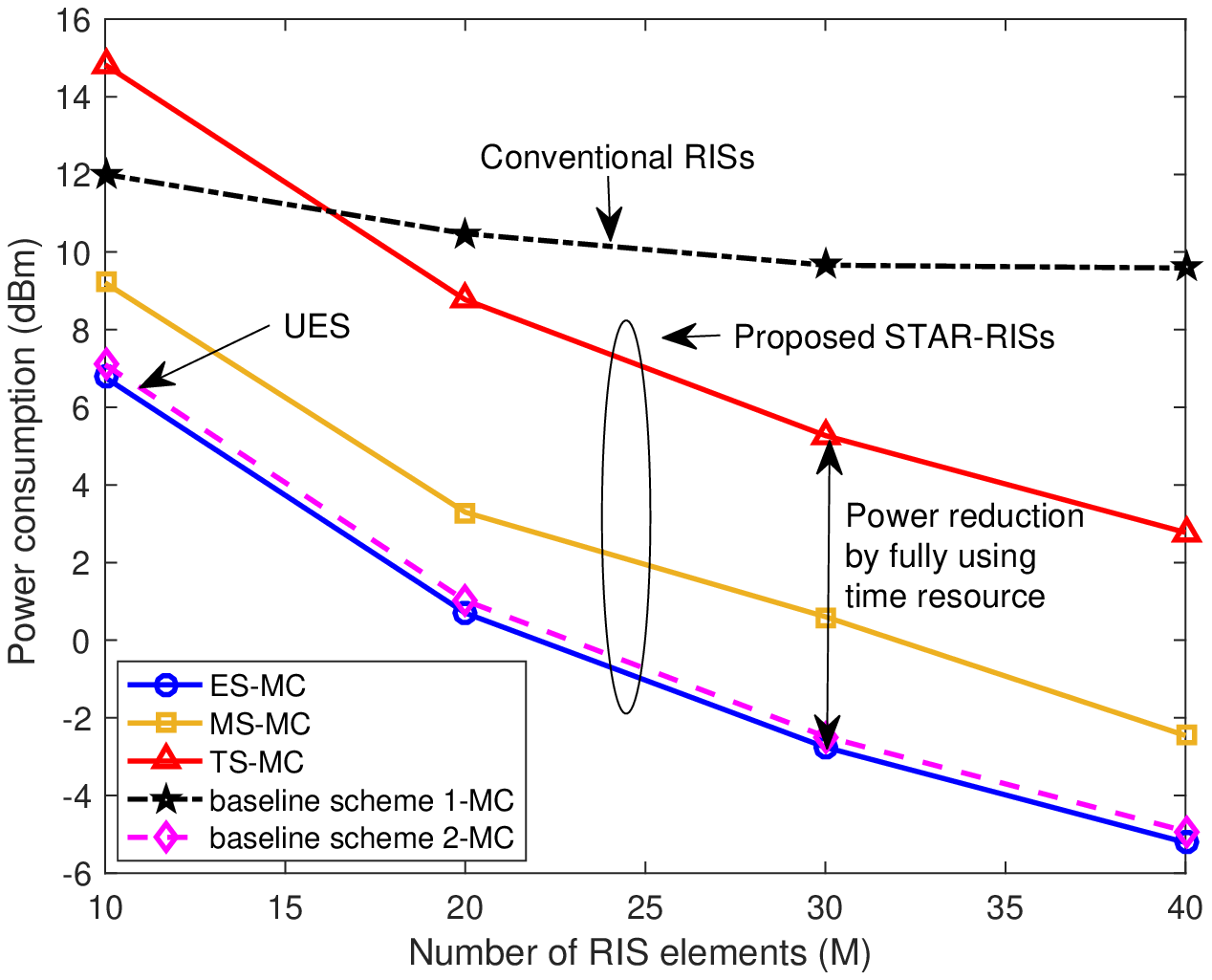}}
\caption{Power consumption versus $M$ for $N=2$.}\label{CvM}
\end{figure}

\vspace{-0.2cm}
In Fig. \ref{CvM}, we investigate the required power consumption versus the number of RIS elements, $M$, for both unicast and multicast transmission. Besides the three proposed operating protocols, the results obtained for the two baseline schemes are provided for comparison. We set $N=2$, $\overline \gamma_0 = 0$ dB for unicast communication, and $N=2$, $\overline \gamma_c = 10$ dB for multicast communication. As can be seen from Fig. \ref{CvM}, the required power consumption for all schemes and scenarios decreases as $M$ increases. This is expected since larger $M$ enable a higher transmission/reflection beamforming gain, which in turn reduces the required power consumption of the BS. Regarding the performance of the three proposed operating protocols for STAR-RISs, TS achieves the best performance for unicast communication, whereas ES is preferable for multicast communication. This can be explained as follows. TS results in interference-free communication for each user, i.e., only one user is to be served in each time instant. This interference-free feature makes TS preferable for unicast communication since it prevents the communication rate of the users from being degraded by inter-user interference. In contrast, for multicast communication, inter-user interference does not exist since the same symbol is sent to all users. Hence, ES becomes appealing since it can make full use of the entire available communication time and allows the users to be served all the time, which is not possible with TS. Moreover, it can also be  observed that ES outperforms MS for both unicast and multicast transmission. This is expected since mathematically MS is a special case of ES.

Regarding the performance comparison between STAR-RISs and baseline scheme 1, as can be observed from Fig. \ref{UCvM}, independent of the adopted operating protocol, STAR-RISs always outperform conventional RISs for unicast communication. The reasons behind this can be explained as follows. Since conventional RISs employ fixed numbers of transmission and reflection elements, they cannot exploit the same DoFs as available to the STAR-RISs to enhance the desired signal strength and mitigate inter-user interference. Therefore, conventional RISs always suffer from the worst performance for unicast communication. However, in Fig. \ref{MCvM}, for multicast communication, we observe that conventional RISs can achieve a higher performance than TS STAR-RIS for small $M$. This is because, compared with TS STAR-RIS, conventional RISs (a special case of MS STAR-RISs) can make full use of the entire available communication time. This benefit allows conventional RISs to achieve a higher performance than TS when $M$ is small, i.e., when the available DoFs are limited and using the entire available communication time dominates the achieved performance. However, when $M$ increases, as can be observed from Fig. \ref{MCvM}, conventional RISs become the worst option again due to the significant loss of DoFs caused by their inflexibility in choosing between transmission and reflection. This limitation also causes the performance gap between conventional RISs and STAR-RISs to become more pronounced as $M$ increases. The above performance comparison confirms the effectiveness of employing STAR-RIS in wireless communication systems.

For baseline scheme 2, a special case of ES, it is interesting to observe that there is a noticeable performance gap between ES and UES for unicast communication in Fig. \ref{UCvM}. However, the performance gap is negligible for multicast communication in Fig. \ref{MCvM}. Recall that the difference between ES and UES only lies in the element-wise and group-wise transmission and reflection amplitude control. The extra DoFs provided by the element-wise amplitude control allow ES to achieve an improved desired signal enhancement and inter-user interference mitigation compared to UES for unicast communication, thus achieving higher performance. For multicast communication, since there is no inter-user interference to be mitigated, the performance gain caused by the element-wise amplitude control vanishes. On the one hand, this result underscores the importance of element-wise transmission and reflection amplitude control for mitigating inter-user interference for unicast communication. On the other hand, the result also implies that UES is a promising operating protocol for multicast communication, where it achieves similar performance as ES. However, compared with ES, the group-wise amplitude control of UES reduces the overhead caused by exchanging configuration information between the BS and the STAR-RIS, which is a major bottleneck for the practical implementation of RISs.
\vspace{-0.6cm}
\subsection{Required Power Consumption Versus Number of BS Antennas}
\begin{figure}[t!]
\centering
\subfigure[Unicast communication, $\overline \gamma _0$ = 0 dB.]{\label{UCvN}
\includegraphics[width= 3in]{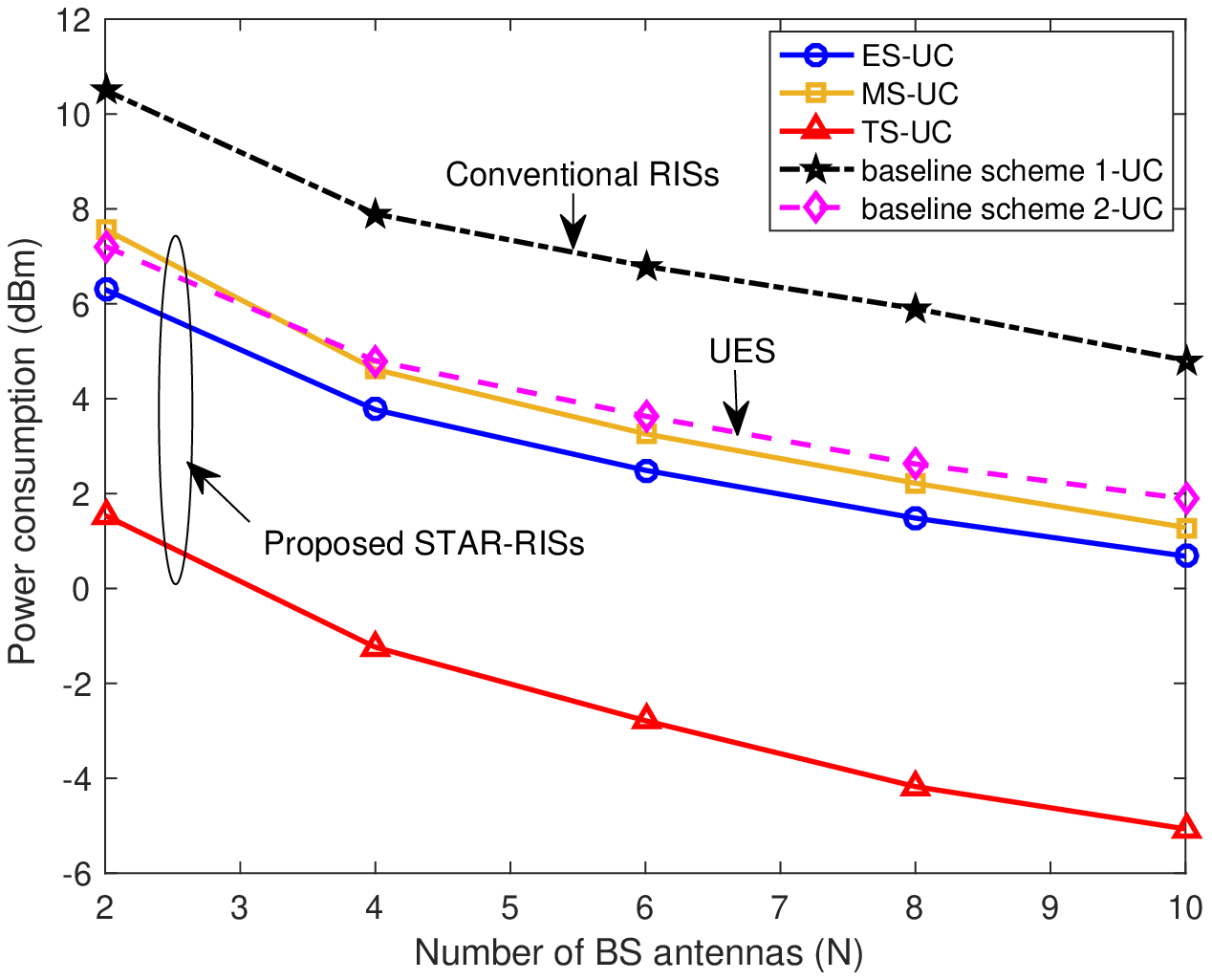}}
\subfigure[Multicast communication, $\overline \gamma _c$ = 10 dB.]{\label{MCvN}
\includegraphics[width= 3in]{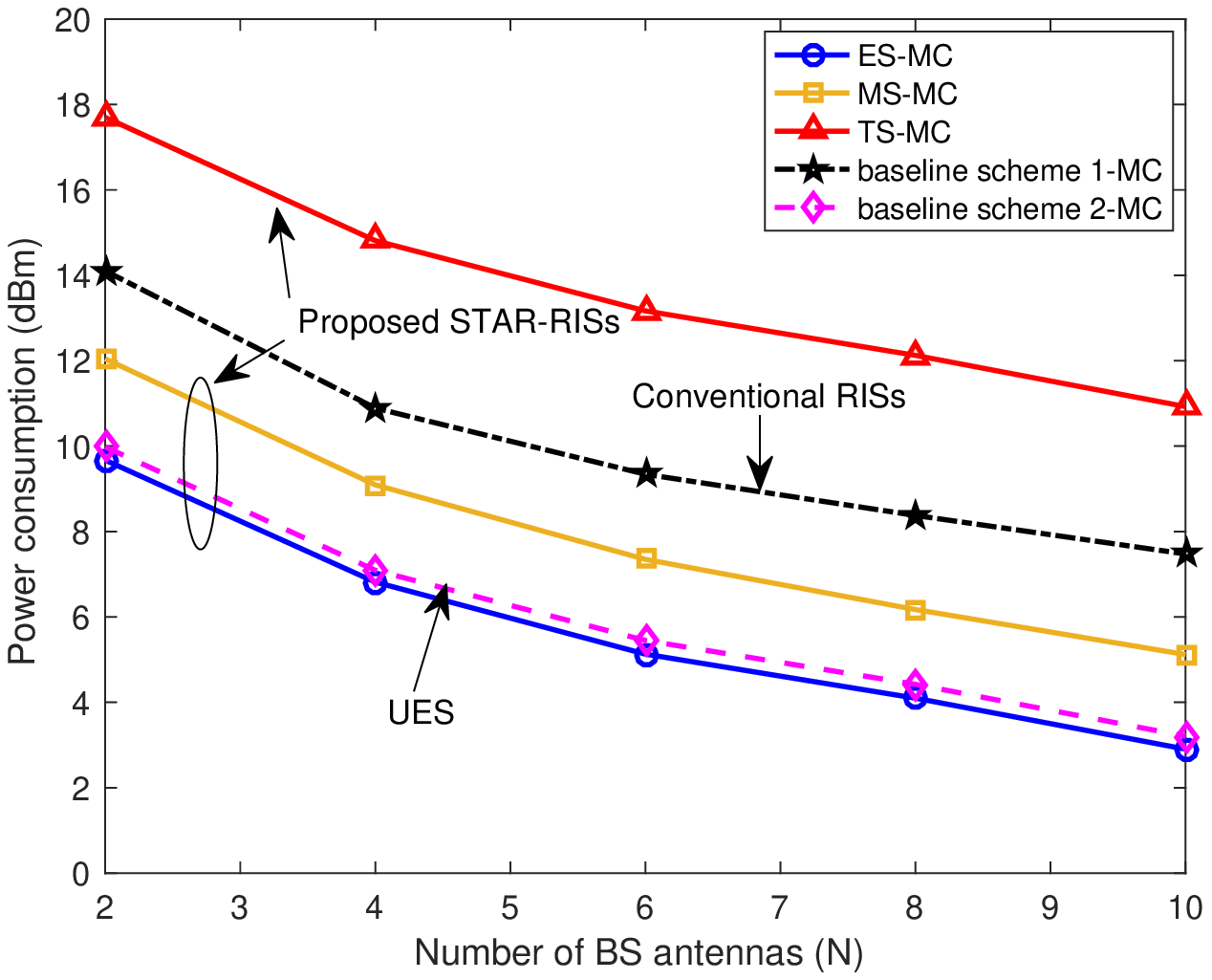}}
\caption{Power consumption versus $N$ for $M=10$.}\label{CvN}
\end{figure}

\vspace{-0.2cm}
In Fig. \ref{CvN}, we study the required power consumption versus the number of BS antennas, $N$, for the proposed operating protocols for STAR-RISs and the baselines for both unicast and multicast communication. We set $M=10$, $\overline \gamma_0 = 0$ dB for unicast communication, and $M=10$, $\overline \gamma_c = 10$ dB for multicast communication. Fig. \ref{CvN} shows that the required power consumption for all schemes decreases as $N$ increases thanks to a higher active beamforming gain. For unicast communication in Fig. \ref{UCvN}, the proposed STAR-RISs outperform conventional RISs since more DoFs for transmission and reflection design can be exploited. As we consider a small $M$, for multicast communication in Fig. \ref{MCvN}, the loss in DoFs for employing conventional RISs is not significant. Thus, in this case, conventional RISs outperform the TS STAR-RIS because of the more efficient exploitation of the time resources. As can be observed in Fig. \ref{UCvN}, there is a considerable performance gap between UES and ES for unicast communication, while UES and ES achieve a similar performance for multicast communication in Fig. \ref{MCvN}. This is consistent with the results in Fig. \ref{CvM}.
\vspace{-0.6cm}
\subsection{Required Power Consumption Versus QoS Requirements}%, when  increases from 0 dB to 25 dB,
\vspace{-0.2cm}
\begin{figure}[!t]
\centering
\subfigure[Unicast communication.]{\label{UCvR}
\includegraphics[width= 3in]{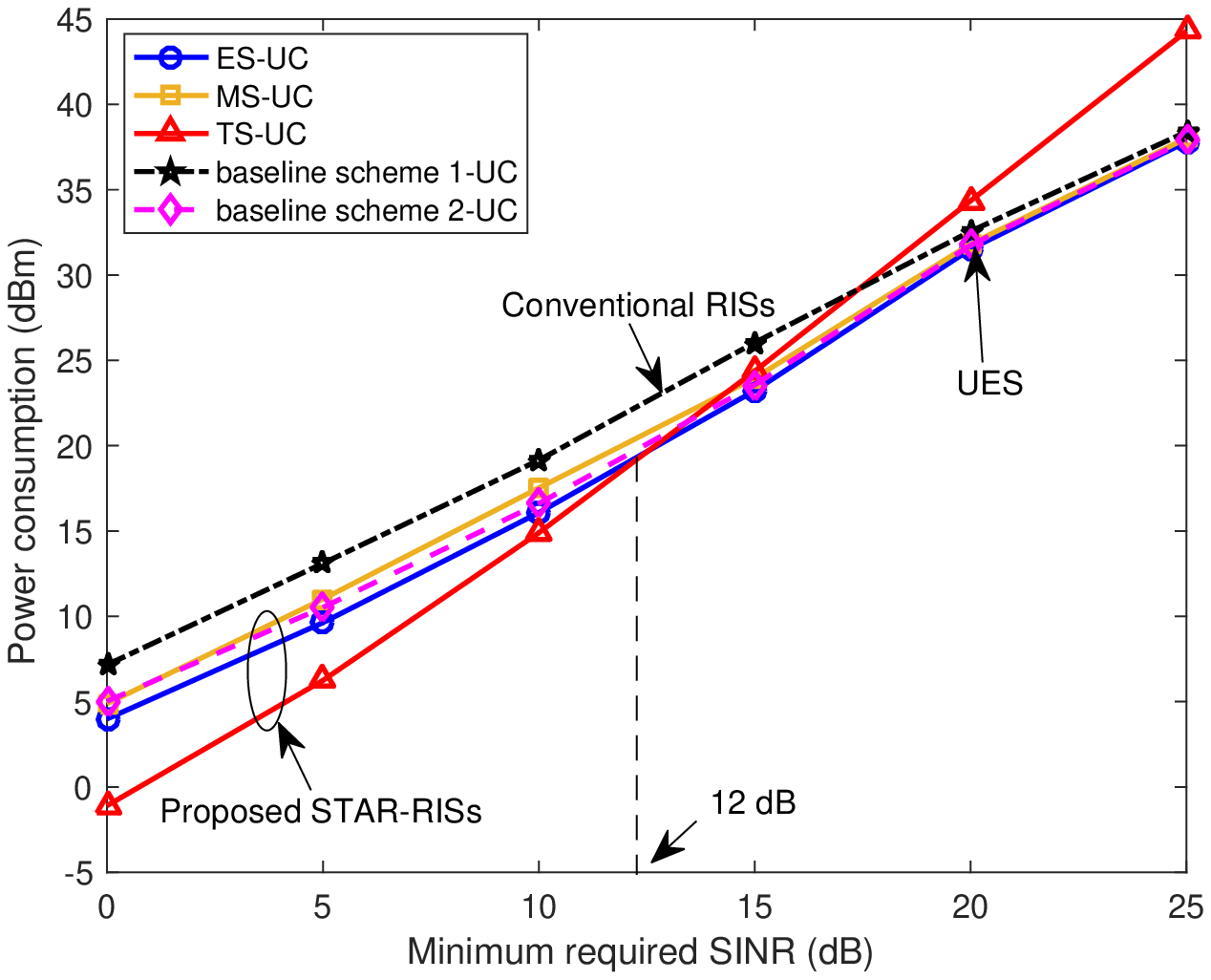}}
\subfigure[Multicast communication.]{\label{MCvR}
\includegraphics[width= 3in]{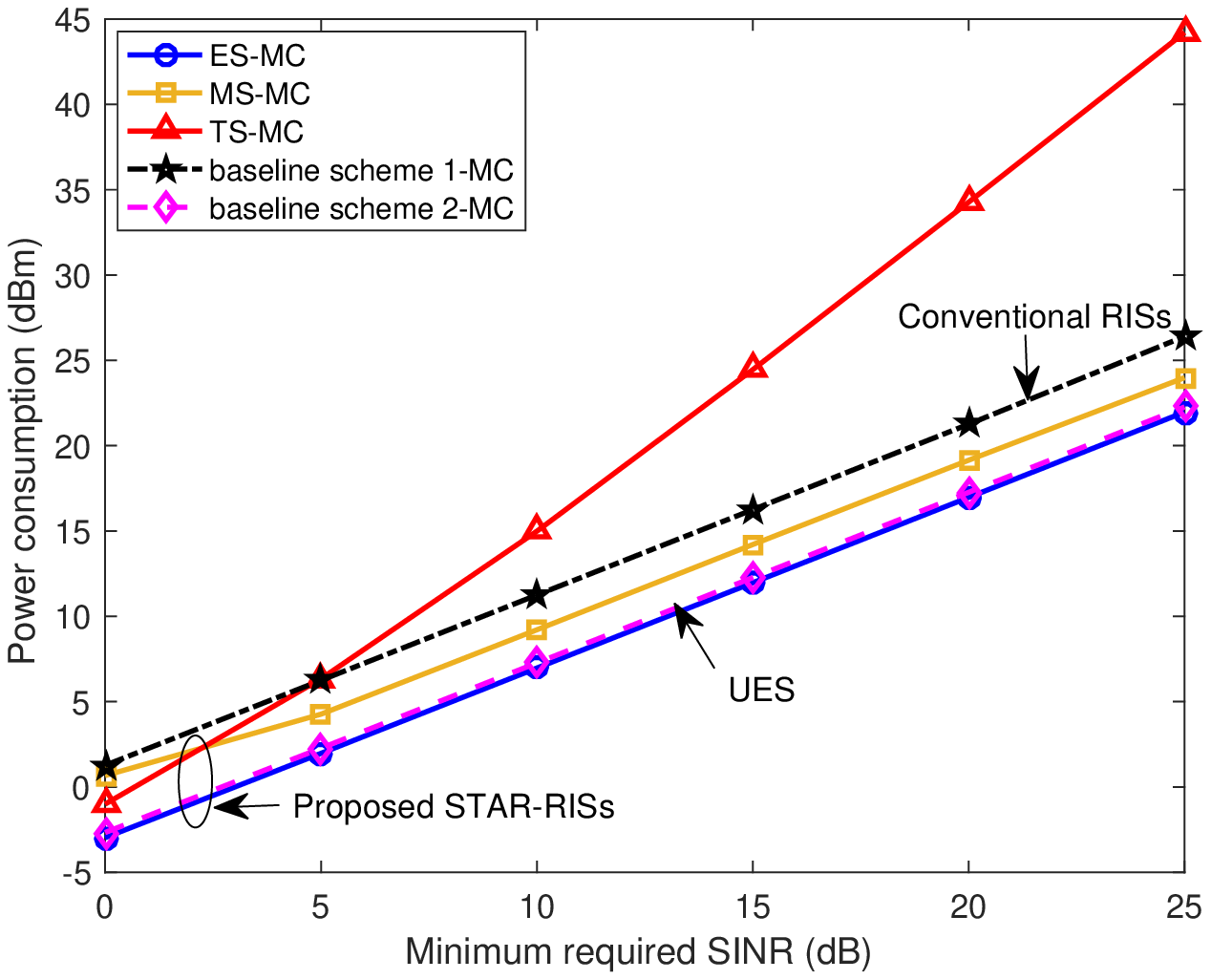}}
\caption{Power consumption versus $\overline \gamma _0$ and $\overline \gamma_c$ for $N=2$ and $M=10$.}\label{CvR}
\end{figure}
In Fig. \ref{CvR}, we study the required power consumption versus the minimum required SINR, $\overline \gamma _0$ and $\overline \gamma _c$, for both unicast and multicast transmission. We set $N=2$ and $M=10$ for all considered schemes. For unicast communication in Fig. \ref{UCvR}, we observe that TS achieves the best performance for small $\overline \gamma _0$ (i.e., when $\overline \gamma _0<12$ dB) but is outperformed by the other schemes for large $\overline \gamma _0$. This is because the interference-free communication for TS is achieved at the expense of an inefficient use of the communication time, which becomes a major performance bottleneck for large $\overline \gamma _0$. In contrast, since for ES, MS, and the two baseline schemes, both users are served during the entire communication time, they achieve a higher performance than TS for large $\overline \gamma _0$ despite the inter-user interference. Similarly, for multicast communication, we can observe from Fig. \ref{MCvR} that the performance gap between TS and the other schemes becomes more pronounced as $\overline \gamma _c$ increases due to the inefficient use of the communication time for TS. The obtained results highlight the importance of employing different operating protocols for different communication objectives and scenarios.
\vspace{-0.2cm}
\section{Conclusions}
In this paper, wireless communication system design for a novel RIS concept enabling simultaneous transmission and reflection was investigated. Based on the basic signal model for the individual RIS elements, three practical operating protocols for STAR-RISs were proposed and their respective advantages and disadvantages were discussed. For each of these operating protocols, the joint active and passive beamforming optimization problem was formulated for minimization of the power consumption of the BS while satisfying the QoS requirements of the users for both unicast and multicast transmission. For the ES and MS protocols, the resulting non-convex problems were efficiently solved by penalty-based iterative algorithms. For the TS protocol, the optimization problem was solved using state-of-the-art algorithms and convex optimization techniques. Numerical results showed that STAR-RISs can significantly reduce the BS power consumption compared to conventional reflecting/transmitting-only RISs. Furthermore, the obtained results also revealed that the TS protocol is preferable for unicast communication and low QoS requirements. However, for unicast communication and high QoS requirements, and for multicast communication, the ES protocol is the best option. These insights provide useful guidelines for the design of STAR-RIS aided wireless communication systems.\\
\indent The results obtained in this paper confirm the effectiveness of employing STAR-RISs for improving the performance of wireless networks, which motivates related future research on STAR-RISs, such as channel estimation and deployment design. More specifically, although the channel acquisition methods proposed for conventional RISs can be applied for STAR-RIS with TS, considerable pilot overhead is required to \emph{consecutively} estimate the transmission and reflection channels. To tackle this issue, channel acquisition methods based on the ES protocol have to be developed to \emph{simultaneously} estimate both channels with less pilot overhead. Moreover, the full-space coverage created by STAR-RISs also imposes new challenges for deployment. Different from conventional reflecting-only RISs, whose optimal deployment strategy requires only to assign either the users or the BSs to its local reflection region~\cite{9326394,Mu_deployment}, the deployment locations and orientations of STAR-RISs have to be further optimized to balance the numbers users on both sides of the STAR-RIS, i.e., the number of T users and the number of R users.
\vspace{-0.2cm}
\section*{Appendix~A: Proof of Theorem~\ref{rank one relax}} \label{Appendix:B}
The relaxed version of problem \eqref{P1 GES2} without rank-one constraint \eqref{Rank W GES} is jointly concave with respect to $\left\{ {{{\mathbf{W}}_k},{{\mathbf{Q}}_k^{\rm{ES}}},{{\bm{\beta}} _k}} \right\}$ and satisfies Slater's constraint qualification~\cite{convex}. Therefore, strong duality holds and the Lagrangian function is given by
\vspace{-0.3cm}
\begin{align}\label{Lagrangian}
\begin{gathered}
  {{\mathcal{L}}} = \sum\nolimits_{k \in \left\{ {t,r} \right\}} {{\rm{Tr}}\left( {{{\mathbf{W}}_k}} \right)} \hfill \\
    + \sum\nolimits_{k \in \left\{ {t,r} \right\}} {{\mu _k}} \left( {\frac{{{{\overline \gamma }_k}}}{2}\left\| {{{\mathbf{Q}}_k^{\rm{ES}}} + {{\mathbf{H}}_k}{{\mathbf{W}}_{\overline k}}{\mathbf{H}}_k^H} \right\|_F^2 - {{\overline \gamma }_k}{\rm{Tr}}\left( {{{\left( {{\mathbf{H}}_k^H{{\mathbf{H}}_k}{\mathbf{W}}_{\overline k}^{\left( n \right)}{\mathbf{H}}_k^H{{\mathbf{H}}_k}} \right)}^H}{{\mathbf{W}}_{\overline k}}} \right)} \right. \hfill \\
   + \left. {\frac{1}{2}\left\| {{{\mathbf{Q}}_k^{\rm{ES}}} - {{\mathbf{H}}_k}{{\mathbf{W}}_k}{\mathbf{H}}_k^H} \right\|_F^2 - {\rm{Tr}}\left( {{{\left( {{\mathbf{H}}_k^H{{\mathbf{H}}_k}{\mathbf{W}}_k^{\left( n \right)}{\mathbf{H}}_k^H{{\mathbf{H}}_k}} \right)}^H}{{\mathbf{W}}_k}} \right)} \right) - \sum\nolimits_{k \in \left\{ {t,r} \right\}} {{\rm{Tr}}\left( {{{\mathbf{Y}}_k}{{\mathbf{W}}_k}} \right)}  + \tau,  \hfill \\
\end{gathered}
\end{align}
\vspace{-0.8cm}

\noindent where $\tau$ is the collection of all terms which do not depend on $\left\{ {{{\mathbf{W}}_k}} \right\}$, and ${{\mu _k}}$ and ${{{\mathbf{Y}}_k}}$ are the Lagrange multipliers associated with constraints \eqref{QoS GES2} and \eqref{H Q W GES}, respectively. Based on the Karush-Kuhn-Tucker (KKT) conditions with respect to ${{{\mathbf{W}}_k}}$, the structure of the optimal ${\mathbf{W}}_k^*$ can be characterized as follows:
\vspace{-0.4cm}
\begin{subequations}\label{KKT}
\begin{align}
\label{KKT1}&\mu _k^* \ge 0,{\mathbf{Y}}_k^* \succeq {\mathbf{0}},\\
\label{KKT2}&{\mathbf{Y}}_k^*{\mathbf{W}}_k^* = {\mathbf{0}},\\
\label{KKT3}&{\nabla _{{\mathbf{W}}_k^*}}{{\mathcal{L}}} = {\mathbf{0}},
\end{align}
\end{subequations}
\vspace{-1.2cm}

\noindent where $\mu _k^*$ and ${\mathbf{Y}}_k^*$ denote the optimal Lagrange multipliers and ${\nabla _{{\mathbf{W}}_k^*}}{\mathcal{L}}$ is the gradient of ${\mathcal{L}}$ with respect to ${\mathbf{W}}_k^*$. Then, ${\nabla _{{\mathbf{W}}_k^*}}{\mathcal{L}} = 0$ can be further expressed as
\vspace{-0.3cm}
\begin{align}\label{KKT=0}
{\mathbf{Y}}_k^* = {{\mathbf{I}}_N}  - {\mathbf{\Omega }}_k^*,
\end{align}
\vspace{-1.2cm}

\noindent where ${\mathbf{\Omega }}_k^*$ is given by
\vspace{-0.4cm}
\begin{align}\label{Gamma}
\begin{gathered}
  {\mathbf{\Omega }}_k^* = {\mu _k}\left( {{{\mathbf{H}}_k}{{\mathbf{Q}}_k^{\rm{ES}}}{\mathbf{H}}_k^H - {{\mathbf{H}}_k}{{\mathbf{H}}_k}{{\mathbf{W}}_k}{\mathbf{H}}_k^H{\mathbf{H}}_k^H + {\mathbf{H}}_k^H{{\mathbf{H}}_k}{\mathbf{W}}_k^{\left( n \right)}{\mathbf{H}}_k^H{{\mathbf{H}}_k}} \right) \hfill \\
   - {\mu _{\overline k}}{{\overline \gamma }_{\overline k}}\left( {{{\mathbf{H}}_{\overline k}}{{\mathbf{Q}}_k^{\rm{ES}}}{\mathbf{H}}_{\overline k }^H + {{\mathbf{H}}_{\overline k}}{{\mathbf{H}}_{\overline k}}{{\mathbf{W}}_k}{\mathbf{H}}_{\overline k }^H{\mathbf{H}}_{\overline k }^H - {\mathbf{H}}_{\overline k }^H{{\mathbf{H}}_{\overline k}}{\mathbf{W}}_k^{\left( n \right)}{\mathbf{H}}_{\overline k }^H{{\mathbf{H}}_{\overline k}}} \right) \hfill \\
\end{gathered}
\end{align}
\vspace{-0.6cm}

\noindent By exploiting the results in [17, Appendix A], it can be proved that ${\rm{Rank}}\left( {{\mathbf{Y}}_k^*} \right) = N - 1$. Furthermore, \eqref{KKT2} implies that ${\rm{Rank}}\left( {{\mathbf{Y}}_k^*} \right) + {\rm{Rank}}\left( {{\mathbf{W}}_k^*} \right) \le N$. Therefore, ${\rm{Rank}}\left( {{\mathbf{W}}_k^*} \right) \le 1$ has to hold. Thus, due to the QoS constraint in \eqref{QoS GES2}, ${\rm{Rank}}\left( {{\mathbf{W}}_k^*} \right) = 1$ for the optimal solution. This completes the proof.
\bibliographystyle{IEEEtran}
\bibliography{mybib}
 \end{document}